\documentclass{ws-ijmpa}
\usepackage[super,compress]{cite}
\usepackage{graphicx}

\usepackage{float}

\def\be{\begin{equation}}
\def\te{\end{equation}}
\def\ee{\end{equation}}
\def\ba{\begin{eqnarray}}
\def\bea{\begin{eqnarray}}
\def\nn{\nonumber\\}
\def\tea{\end{eqnarray}}
\def\ea{\end{eqnarray}}
\def\eea{\end{eqnarray}}

\begin{document}
\markboth{Cantarutti-Calzetta}{Dissipative type theories for Bjorken and Gubser flows}

\catchline{}{}{}{}{}
%
%%%%%%%%%%%%%%%%%%%%%%%%%%%%%%%%%%%%%%%%%%%%%%%%%%%%%%%%%%%%%%%%%%%%

\title{Dissipative type theories for Bjorken and Gubser flows}

\author{Lucas Cantarutti}

\author{Esteban Calzetta\footnote{calzetta@df.uba.ar}}

\address{Departamento de F\'\i sica, FCEyN-UBA, and IFIBA, UBA-CONICET, Buenos Aires, Argentina}

\maketitle

\begin{history}
\received{Day Month Year}
\revised{Day Month Year}
\end{history}

\begin{abstract}
We use the dissipative type theory (DTT) framework to solve for the evolution of conformal fluids in Bjorken and Gubser flows from isotropic initial conditions. The results compare well with both exact  and other hydrodynamic solutions in the literature. At the same time, DTTs enforce the Second Law of thermodynamics as an exact property of the formalism, at any order in deviations from equilibrium, and are easily generalizable to more complex situations.

\end{abstract}

\ccode{PACS numbers:47.75.+f, 24.10.Nz, 25.75.-q}

%\tableofcontents

\section{Introduction}
The success of hydrodynamics in describing relativistic heavy ion collisions\cite{RR19} and the theoretical conjecture of an absolute lowest limit for viscosity\cite{KSS05} has focused attention of the development of a relativistic hydro and magnetohydro dynamics of viscous fluids\cite{Anile89,RZ13}. While this is a relatively old subject\cite{Catt48,Catt58}, early attempts\cite{LL6,Eck40} have been marred by causality and stability problems\cite{HL83,HL85,HL88a,HL88b,Ols90,HP01,DKKM08,PKR10,GPRR19}. Eventually a number of different formulations arose, such as extended thermodynamics\cite{PJCV80,JP91,MR93,JCVL10}, Israel-Stewart\cite{S71,I72,I76,IS76,IS79a,IS79b,IS80,I88,OH90}, BRSSS\cite{BRSSS08,BHMR08}, DNMR \cite{DMNR10,DMNR11,DMNR11b,DMNR12,DMNR12b,DMNR14,DMNR12c,DMNR14b}, anisotropic hydrodynamics\cite{ST14,ST14b,FMR15,FRST16} and viscous anisotropic hydrodynamics\cite{BHS14,BNR16a,BNR16b,NMR17}. 

Those approaches that stress ensuring nonnegative entropy production along with energy-momentum conservation are particularly relevant to this paper \cite{L08,JBP13,CJPR15}. The difficulty of modelling relativistic viscous flows is compounded by the fact that these flows are liable to become unstable \cite{M94,M07,MM06,SSGT06,MM07,MT07,RSA08,RS10,IRS11,ARS13,MSS16,CK16} or else enter into a turbulent regime \cite{K08,FW11,CR11,F13,AKLN14,ED18}, wherefrom any initially ``small'' perturbation may grow without limit, see also \cite{GNGR18}. The alternative of obtaining a nonperturbative description by actually resumming a perturbative expansion, to the best of our knowledge, has not been carried out except in some simple, highly symmetric flows\cite{DN16,AS16,FRS16}.  

Dissipative type theories (DTTs)\cite{L72,LMR86,GL90,GL91,NR95,RN97,BR99,M99,C98,CT01,PRC09,PRC10a,PRC10c,PRC10b,PRC12,MGC17,LRR18}
 were introduced as a way to provide relativistic and thermodynamic consistency in arbitrary flows independently of any approximations. We believe for this reason alone they deserve to be seriously considered as the proper relativistic generalization of the Navier-Stokes equations. However, these appealing features would not be enough if they cannot pass the few tests we have to evaluate hydrodynamic theories.

Among these, the study of conformal fluids in Bjorken\cite{BJOR83} and Gubser\cite{G10,GY11} flows stands out. Both are highly symmetric flows (to be described in more detail below) where an exact solution of the kinetic theory equations with an Anderson-Witting collision term\cite{AW1,AW2,TI10} is available. These allows for a detailed comparison between the hydrodynamic theory of choice and the exact underlying theory it aims to reproduce. Although the high symmetry of these flows may be misleading, they have provided a highly valuable test bench for relativistic hydrodynamics.

In latter years a number of theories have been tested in these scenarios\cite{MS10b,FRS13a,FRS13b,DHMNS14a,DHMNS14b,HM15,NRS15,TRFS15,HBDMNNRS16,MMcNH17,ChHPV18}
, which have also been used to study hydrodynamic fluctuations\cite{AMT17,MSch18} as well as the hydrodynamization and thermalization processes\cite{BCM18,BCKM19,ChHPV19,ChH19}, but to the best of our knowledge DTTs have not been tried so far. This paper aims to fill this gap, showing that a suitable DTT performs at a level satisfactorily close to the exact solutions in both flows. 

The rest of the paper is organized as follows. In next section we ellaborate on why the validity of the Second law should not be taken for granted in hydrodynamics, even when derived from kinetic theories for which an $H$ theorem may be proven. We also discuss why thermodynamic consistency leads us to DTTs, and describe the kind of DTT to be tested in the remainder of the paper. The following two sections apply this DTT to conformal fluids in Bjorken and Gubser flows. We only compare our results to the exact and third order Eckart theories\cite{ChHPV18, J13}, since detailed comparison to other frameworks may be found in the literature. We conclude with some brief final remarks.

This paper has four appendices. In Appendix A we expand on some properties of the phase space of a relativistic particle which are relevant to our discussion. Appendix B and C are the detail of the relevant tensors calculation in the Bjorken and Gubser flow respectively. Finally, in Appendix D we compare DTTs to the better known so-called ``second order'' hydrodynamic theories, taking references \cite{DMNR10,DMNR11,DMNR11b,DMNR12,DMNR12b,DMNR14,DMNR12c,DMNR14b} and \cite{JRS14,FJMRS15,TJR17,TVNH19} as representative formulations.

\section{From kinetic theories to hydrodynamics}
We consider the evolution of a relativistic, conformally invariant gas in a curved space time described by a metric $g_{\mu\nu}$ with signature $\left( -,+,+,+\right)$. The state of a particle is described by a point $\left(x^{\mu},p_{\mu}\right)$ in phase space, where $x^{\mu}$ denotes a point in the spacetime manifold, and $p_{\mu}$ are the covariant components of a vector in the tangent space at $x$. The particles are massless, so the momentum variables lie on the mass shell $p^2=0$, and have positive energy $p^0\ge 0$. We develop first the kinetic theory description, and then the transition to hydrodynamics.

\subsection{Kinetic theory}
In kinetic theory the gas is described by a one-particle distribution function (1pdf) $f\left( x,p\right) $, which is a nonnegative scalar (see \ref{rps} for further details on the geometry of relativistic phase space) obeying the transport equation 

\be
p^{\mu}\nabla_{\mu}f =I_{col}\left[ f\right]
\label{Boltzmann} 
\te 
where $\nabla$ is the covariant derivative eq. (\ref{nabla}) and the collision integral $I_{col}$ must be specified. For simplicity we assume Maxwell-Boltzmann statistics, the generalization to quantum statistics is immediate. In equilibrium the one-particle distribution function obeys 

\be 
f\equiv f_{eq}=e^{\beta_{\mu}p^{\mu}}
\te 
where $\beta_{\mu}$ is a timelike Killing field: $\beta_{\mu;\nu}+\beta_{\nu;\mu}=0$. Therefore we request 

\be 
I_{col}\left[ f_{eq}\right]\equiv 0
\te 
It is convenient to introduce the temperature $T$ from $\beta_{\mu}=u_{\mu}/T$, with $u^2=-1$. For a general $f$ the energy momentum tensor (EMT)

\be 
T^{\mu\nu}=\int \frac{Dp}{\sqrt{-g}}\; p^{\mu}p^{\nu}f
\label{EMT}
\te 
where $Dp$ is the invariant measure eq. (\ref{invariant}). In equilibrium the EMT adopts the perfect fluid form

\be 
T^{\mu\nu}_{eq}=\epsilon u^{\mu}u^{\nu}+p\Delta^{\mu\nu}
\te 
where the pressure $p=\epsilon /3$, $\Delta^{\mu\nu}=g^{\mu\nu}+u^{\mu}u^{\nu}$ and the energy density 

\be 
\epsilon\equiv\epsilon_{eq} =\int \frac{Dp}{\sqrt{-g}}\; \left( u_{\mu}p^{\mu}\right) ^2f_{eq}=\sigma_{SB}T^4
\te 
where $\sigma_{SB}=3/\pi^2$ is the Stefan-Boltzmann constant. Conservation of the EMT

\be 
T^{\mu\nu}_{;\nu}=0
\label{7}
\te 
imposes a new constraint on the collision integral 

\be 
\int \frac{Dp}{\sqrt{-g}}\; p^{\mu}I_{col}\left[ f\right]\equiv 0
\label{conservation}
\te 
for \emph{any} $f$. We also have the entropy current

\be 
S^{\mu}=\int \frac{Dp}{\sqrt{-g}}\; p^{\mu}f\left[ 1-\ln f\right] 
\label{ktec}
\te
In equilibrium $S^{\mu}=su^{\mu}$, $s\equiv s_{eq}=(4/3)\epsilon_{eq}/T$. The relativistic Second Law reads

\be 
S^{\mu}_{;\mu}\ge 0
\te
Explicitly 

\be 
S^{\mu}_{;\mu}=-\int \frac{Dp}{\sqrt{-g}}\; \ln f\;I_{col}\left[ f\right] 
\label{kentroprod}
\te 
so the Second Law is enforced if the collision integral satisfies the $H$ theorem 

\be 
\int \frac{Dp}{\sqrt{-g}}\; \ln f\;I_{col}\left[ f\right]\le 0
\label{BHT}
\te 
for \emph{any} one-particle distribution function $f$. 

Later on we shall adopt a collision integral of the Anderson-Witting form\cite{AW1,AW2,TI10}

\be 
I_{col}=\frac{U_{\mu}p^{\mu}}{\tau_R}\left[ f-f_{eq}\right] 
\label{AW}
\te 
where $U^{\mu}$ is an unit future oriented timelike vector to be specified, $f_{eq}=\exp\left[ U_{\mu}p^{\mu}/T_{0}\right] $, and the relaxation time $\tau_R$ describes the dissipative effects in the theory. The conservation of the EMT eq. (\ref{7}) becomes 

\be 
T^{\mu}_{\nu}U^{\nu}=-\epsilon_{eq}U^{\mu}
\label{LL}
\te 
Therefore $U^{\mu}$ and $T_{0}$ are derived from $T^{\mu\nu}$ through the Landau-Lifshitz prescription\cite{LL6}, namely $U^{\mu}$ is the timelike eigenvector of the EMT, and $\sigma_{SB}T_{0}^4$ the corresponding eigenvalue. The $H$ theorem follows from the identity 

\be 
\int \frac{Dp}{\sqrt{-g}}\; \ln f_{eq}\;I_{col}\left[ f\right]=\frac{1}{\tau_R}\left[ U_{\mu}U_{\nu} T^{\mu\nu}-\epsilon_{eq}\right] =0
\te 
Because then 

\be 
\int \frac{Dp}{\sqrt{-g}}\; \ln f\;I_{col}\left[ f\right]=\int \frac{Dp}{\sqrt{-g}}\; \ln \left[ \frac f{f_{eq}}\right] \;I_{col}\left[ f\right]\le 0
\te 
and both $U^{\mu}$ and $p^{\mu}$ are timelike and future oriented.

To sustain conformal invariance we must further have the relationship\cite {LRR18}

\be 
T_0\tau_R =c=\;\mathrm{constant}
\label{confinv}
\te

\subsection{Hydrodynamics}
Once $U^{\mu}$ and $T_{0}$ have been identified from eq. (\ref{LL}), we can always write

\be 
T^{\mu\nu}=T_0^{\mu\nu}+\Pi^{\mu\nu}
\label{decomp}
\te 
where 

\be 
T_0^{\mu\nu}=\sigma_{SB}T_0^4\left[ U^{\mu}U^{\nu}+\frac13 h^{\mu\nu}\right] 
\label{idealEMT}
\te 
$h^{\mu\nu}=g^{\mu\nu}+ U^{\mu}U^{\nu}$. $\Pi^{\mu\nu}$ is the so-called viscous EMT 

\be 
\Pi^{\mu\nu}= H^{\mu\nu}_{\rho\sigma}T^{\rho\sigma}=\int \frac{Dp}{\sqrt{-g}}\; H^{\mu\nu}_{\rho\sigma} p^{\rho}p^{\sigma}\;f
\label{viscousEMT}
\te 

\be 
 H^{\mu\nu}_{\rho\sigma}=\frac12\left[ h^{\mu}_{\rho}h^{\nu}_{\sigma}+h^{\mu}_{\sigma}h^{\nu}_{\rho}-\frac23h^{\mu\nu}h_{\rho\sigma}\right] 
\label{hproj}
\te
The conservation equations (\ref{7}) become
 
\bea 
\dot \epsilon +\frac43\epsilon U^{\nu}_{;\nu}+\Pi^{\nu\rho}U_{\nu;\rho}&=&0\nn
\frac13h^{\mu\nu}\epsilon_{,\nu}+\frac43\epsilon \dot U^{\mu}+h^{\mu}_{\nu}\Pi^{\nu\rho}_{;\rho}&=&0
\tea 
$\dot X=U^{\mu}X_{;\mu}$. The task of hydrodynamics is to close these equations by either providing constitutive relations which define $ \Pi^{\nu\rho}$ as a functional of $U^{\mu}$ and $T_0$, or else by adding supplementary equations. The first strategy has led to the so-called \emph{first order} theories\cite{LL6,Eck40}. Although they may be workable in some cases, in general they have causality and stability problems\cite{HL83,HL85,HL88a,HL88b,Ols90,HP01,DKKM08,PKR10,GPRR19}. We shall explore the second strategy. 

The idea is to consider a restricted class of 1pdfs $\mathbf{f}\left[x,p;\zeta^n\left( x\right) \right] $, parametrized in terms of a finite number of position dependent hydrodynamical variables $\zeta^{n}$, $n=1,\ldots N$. We shall consider the case where $\zeta^{n}=\zeta^{\mu_1\ldots\mu_n}$ is a totally symmetric tensor, traceless on any pair of indexes. They include but are not restricted to $\zeta^1=\beta^{\mu}=U^{\mu}/T$, where $T$ is a dimensionful variable which in equilibrium agrees with $T_0$.  

The parametrized one particle distribution function will not be a solution of the Boltzmann equation (\ref{Boltzmann}). Instead we choose a set of $N$ functions\cite{DMNR12,DMNR12b} $R_n\left(x^{\mu},p_{\mu}\right)p_{\mu_1}\ldots p_{\mu_n}$, where the $R_n$ are scalars, and request the momentum equations

\be 
\int \frac{Dp}{\sqrt{-g}}\; R_np_{\mu_1}\ldots p_{\mu_n}\left\{p^{\mu}\nabla_{\mu}\mathbf{f}\left[p;\zeta^n \right]  -I_{col}\left[ \mathbf{f}\right]\right\rbrace =0
\label{hydroeqs}
\te 
which reduce to (see \ref{rps})

\be 
 A^{\mu}_{\mu_1\ldots\mu_n;\mu}-K_{\mu_1\ldots\mu_n}=I_{\mu_1\ldots\mu_n}
\label{hydrocons}
\te 
where

\bea
A^{\mu}_{\mu_1\ldots\mu_n}&=&\int \frac{Dp}{\sqrt{-g}}\; R_np_{\mu_1}\ldots p_{\mu_n}p^\mu\mathbf{f}\nn
K_{\mu_1\ldots\mu_n}&=&\int \frac{Dp}{\sqrt{-g}}\; p_{\mu_1}\ldots p_{\mu_n}\left(p^{\mu}\nabla_{\mu}R_n\right)\mathbf{f}\nn
I_{\mu_1\ldots\mu_n}&=&\int \frac{Dp}{\sqrt{-g}}\;  R_np_{\mu_1}\ldots p_{\mu_n}I_{col}\left[ \mathbf{f}\right]
\label{hydroconrels}
\tea

\subsection{From the Second Law to DTTs} 
Let us now consider how enforcing the Second Law constrains the above scheme. 

It is natural to assume that the hydrodynamic entropy current is just the restriction of the kinetic theory current eq. (\ref{ktec}) to the class of parameterized one particle distribution functions 

\be 
S^{\mu}=\int \frac{Dp}{\sqrt{-g}}\; p^{\mu}\mathbf{f}\left[ 1-\ln \mathbf{f}\right] 
\label{htec}
\te
Then we obtain the entropy production

\be 
S^{\mu}_{;\mu}=-\int \frac{Dp}{\sqrt{-g}}\; \ln \mathbf{f}\;p^{\mu}\nabla_{\mu} \mathbf{f}
\label{entroprod}
\te 
The problem is that we cannot bring the $H$ theorem to bear, because $\mathbf{f}$ is \emph{not} a solution of eq. (\ref{Boltzmann}). Although it is possible to proceed on a case by case basis, it should be clear that if we want positive entropy production to follow directly from the hydrodynamic equations (\ref{hydrocons}) alone, then we must link eqs (\ref{hydroeqs}) and (\ref{entroprod}) by assuming

\be
\ln \mathbf{f}=\sum_nh^{\left(n\right)\mu_1\ldots\mu_n}\left[\zeta^1,\zeta^2,\ldots\right]\mathcal{R}_np_{\mu_1}\ldots p_{\mu_n}
\te
since then it follows that (see  \ref{rps})

\be 
S^{\mu}_{;\mu}=-\int \frac{Dp}{\sqrt{-g}}\; \ln \mathbf{f}\;I_{col}\left[ \mathbf{f}\right] \ge 0
\te 
from the $H$ theorem. Since now $\mathbf{f}$ depends on the $\zeta^n$ parameters only through the $h^{\left(n\right)}$ tensors, it is further natural to identify them, and we get as our ansatz for the 1pdf

\be
f_{DTT}=\exp\left\{\sum_nR_n\zeta^{\mu_1\ldots\mu_n}p_{\mu_1}\ldots p_{\mu_n}\right\}
\label{fDTT}
\te
The currents $A^{\mu}_{\mu_1\ldots\mu_n}$ derive from a Massieu function current

\be
A^{\mu}_{\mu_1\ldots\mu_n}=\frac{\partial\Phi^{\mu}}{\partial \zeta^{\mu_1\ldots\mu_n}}
\label{gen}
\te 
where

\be 
\Phi^{\mu}=\int \frac{Dp}{\sqrt{-g}}\; p^{\mu}f_{DTT}
\label{gen2}
\te
If we have chosen $\beta_{\mu}$ as one of the hydrodynamic variables, and $p^{\mu}$ as the corresponding function of momentum, then 

\be
\Phi^{\mu}=\frac{\partial\Phi}{\partial \beta_{\mu}}
\te

\be 
\Phi=\int \frac{Dp}{\sqrt{-g}}\; f_{DTT}
\label{genphi}
\te
The entropy current now reads

\be 
S^{\mu}=\Phi^{\mu}-\sum_n\zeta^{\mu_1\ldots\mu_n}A^{\mu}_{\mu_1\ldots\mu_n}
\label{Massieu}
\te 
where $A^{\mu}_{\mu_1}=T^{\mu}_{\mu_1}$ is the EMT, and (see \ref{rps})

\be 
S^{\mu}_{;\mu}=-\sum_n\zeta^{\mu_1\ldots\mu_n}I_{\mu_1\ldots\mu_n}
\te 
so we may state the $H$ theorem in purely hydrodynamic terms as

\be 
\sum_n\zeta^{\mu_1\ldots\mu_n}I_{\mu_1\ldots\mu_n}\le 0
\label{dtth}
\te
The converse is also true\cite{L72}: if positive entropy production must follow from a set of conservation laws (\ref{hydrocons}), then there must be a linear relationship

\be 
S^{\mu}_{;\mu}=-\sum_n\zeta^{\mu_1\ldots\mu_n}\left[ A^{\mu}_{\mu_1\ldots\mu_n;\mu}-K_{\mu_1\ldots\mu_n}\right]
\te 
for some parameters $\zeta^{\mu_1\ldots\mu_n}$ such that the $H$ theorem eq. (\ref{dtth}) holds. But then there must be a  Massieu current which is the generating vector for the currents, as in eq. (\ref{gen}), and the entropy current takes the form eq. (\ref{Massieu}). Either way we are led to adopt a DTT scheme.

\subsection{DTTs and entropy production}
The analysis so far shows that enforcing the Second Law within a hydrodynamical framework naturally suggests a DTT approach, but offers little guidance on how to choose the hydrodynamical parameters $\zeta^n$ and their conjugated functions of momentum. The entropy production variational method (EPVM)\cite{PRC10a,PRC13a,Cal13a} may be called upon to fill this gap. 

The idea is that the best ansatz for the parameterized one particle distribution function is the one that is an extreme of entropy production eq. (\ref{kentroprod}) for a given EMT eq. (\ref{EMT}). Enforcing this last constraint through Lagrange multipliers $\lambda_{\mu\nu}$ we obtain the variational principle

\be
\frac{\delta S}{\delta f\left(x,p\right)}=0
\te
where

\be 
S=-\int \frac{Dp}{\sqrt{-g}}\; \left[\ln f\;I_{col}\left[ f\right] +\lambda_{\mu\nu}p^{\mu}p^{\nu}f\right]
\te
For concreteness, let us assume an Anderson-Witting collision integral eq. (\ref{AW}). Since in the end we want variations that leave $T^{\mu\nu}$ fixed, they will not change $U^{\mu}$ and $T_0$ either. It is simplest to consider only variations that leave $U^{\mu}$ and $T_0$ unchanged, so that 

\be
\int \frac{Dp}{\sqrt{-g}}\; U_{\mu}p^{\mu}p^{\rho}\delta f=0
\label{constraints}
\te
So we get the variational equation

\be
\int \frac{Dp}{\sqrt{-g}}\; \left\{\frac{U_{\mu}p^{\mu}}{\tau_R}\left[ 1-\frac{f_{eq}}f+\ln \left(f/f_{eq}\right)\right] +\lambda_{\mu\nu}p^{\mu}p^{\nu}\right\}\delta f\left(x,p\right)=0
\label{vareq}
\te
Because of eq. (\ref{constraints}) and the mass shell condition we may assume $\lambda_{\mu\nu}U^{\nu}=\lambda^{\mu}_{\mu}=0$. It is clear that when $\lambda_{\mu\nu}=0$ the solution is $f=f_{eq}$. The general solution to the variational problem takes the DTT form eq. (\ref{fDTT}) when $\lambda_{\mu\nu}$ is small. If we write

\be
f=e^{\beta_{0\mu}p^{\mu}+z}
\te
$\beta_0^{\mu}=U^{\mu}/T_0$, then to first order in $\lambda_{\mu\nu}$ we get

\be
z=\frac{\tau_R}2\lambda_{\mu\nu}\frac{p^{\mu}p^{\nu}}{\left(-U_{\rho}p^{\rho}\right)}+\delta\beta_{\mu}p^{\mu}
\te
The last term is a necessary shift to enforce eq. (\ref{constraints}); it is best not to compute it explicitly, but simply enforce the Landau-Lifshitz prescription at the hydrodynamical level. Defining $\beta_{\mu}=u_{\mu}/T=\beta_{0\mu}+\delta\beta_{\mu}$ we get the one particle distribution function

\be
f_{DTT}=e^{\beta_{\mu}p^{\mu}+\left(\zeta_{\mu\nu}/T\right)p^{\mu}p^{\nu}/\left(-U_{\rho}p^{\rho}\right)}
\label{betazeta}
\te
where we have defined $\zeta_{\mu\nu}/T=\tau_R\lambda_{\mu\nu}/2$. This is a DTT with hydrodynamical variables $\beta_{\mu}$ and $\zeta_{\mu\nu}/T$ and conjugated functions $p^{\mu}$ and $p^{\mu}p^{\nu}/\left(-U_{\rho}p^{\rho}\right)$. Observe that we have the constraints that $U^{\mu}$ is the Landau-Lifshitz velocity of the fluid and that $\zeta_{\mu\nu}U^{\nu}=\zeta^{\mu}_{\mu}=0$. These constraints must be enforced \emph{after} the currents are derived from the generating vector. Also, since not all of the components of $\zeta^{\mu\nu}$ are independent, we only enforce a subset of the conservation laws (\ref{hydrocons}). Namely, we only enforce the traceless, transverse part of the conservation law for $A^{\mu}_{\mu_1\mu_2}$. 

Concretely, we obtain the hydrodynamical equations

\bea
T^{\mu\nu}_{;\nu}&=&0\nn
H^{\mu\nu}_{\rho\sigma}\left[A^{\rho\sigma\tau}_{;\tau}-K^{\rho\sigma}-I^{\rho\sigma}\right]&=&0
\label{DynamicEqs}
\tea
($H^{\mu\nu}_{\rho\sigma}$ is defined in eq. (\ref{hproj})), where

\bea
T^{\mu\nu}&=&\int \frac{Dp}{\sqrt{-g}}\;p^{\mu}p^{\nu}f_{DTT}\nn
A^{\mu\nu\tau}&=&\int \frac{Dp}{\sqrt{-g}}\;p^{\tau}\frac{p^{\mu}p^{\nu}}{\left(-U_{\rho}p^{\rho}\right)}f_{DTT}\nn
K^{\rho\sigma}&=&\int \frac{Dp}{\sqrt{-g}}\;p^{\rho}p^{\sigma}\left(p^{\mu}\nabla_{\mu}\left(-U_{\rho}p^{\rho}\right)^{-1}\right)f_{DTT}\nn
I^{\mu\nu}&=&\int \frac{Dp}{\sqrt{-g}}\;\frac{p^{\mu}p^{\nu}}{\left(-U_{\rho}p^{\rho}\right)}I_{col}\left[f_{DTT}\right]
\label{DTTeqs}
\tea 
The $H$ theorem reads $\zeta_{\mu\nu}I^{\mu\nu}\le 0$ and it is a direct consequence of the kinetic theory $H$ theorem eq. (\ref{BHT}).

The resulting theory is close to the so-called anisotropic hydrodynamics\cite{ST14,ST14b,FMR15,FRST16}, which is based on the ansatz

\be
f_{AH}=e^{-\left[\left(U_{\mu}p^{\mu}/T\right)^2-2\left(\zeta_{\mu\nu}p^{\mu}p^{\nu}/T\right)\right]^{1/2}}
\label{AHf}
\te 
The equations of motion are EMT conservation and an equation for particle number, and the Second Law holds. Indeed our DTT could be seen as an approximation to anisotropic hydrodynamics when the departure from isotropy is small. In spite of this ``approximation'',  the Second Law is nevertheless rigorously enforced in the DTT.

If we further expand $f_{DTT}$ to first order in $\zeta_{\mu\nu}$ we obtain the Grad approximation to hydrodynamics\cite{Grad42,Grad49}.

The DTT we have developed is different from the so-called ``statistical'' DTTs\cite{NR95}, which are based on the ansatz

\be
f_{sDTT}=e^{\beta_{0\mu}p^{\mu}+\bar\zeta_{\mu\nu}p^{\mu}p^{\nu}}
\te
For further discussion of statistical DTTs see refs. \citen{cal15,AC17}.

\section{Bjorken flow}
In this section we shall use our DTT (\ref{DynamicEqs}), with the constitutive relations eqs. (\ref{DTTeqs}), to study Bjorken flow.

Bjorken flow is the first qualitatively successful hydrodynamic description of a relativistic heavy ion collision. It describes the collision of two infinitely thin slabs of matter of infinite spatial extension, moving towards each other at the speed of light. In spite of its simplicity it yields concrete predictions, such as a rapidity plateau and, more generally, the so-called Baked Alaska scenario\cite{BKT93}. 

Bjorken flow is commonly expressed in Milne coordinates $x^\mu=(\tau,x,y,\eta)$, where
\be
\tau=\sqrt{t^2-z^2},\;\;\;\;\eta=\tanh^{-1}(z/t)
\ee
The line element is 
\be
ds^2=-d\tau^2+dx^2+dy^2+\tau^2d\eta^2
\ee
and the nontrivial Christoffel symbols are
\bea
\Gamma^\tau_{\eta\eta}=\tau,\;\;\;\;\Gamma^\eta_{\tau\eta}=1/\tau
\tea
The 4-velocity of the flow is defined as $u^{\mu}=U^{\mu}=(1,0,0,0)$ with the normalization $u^\mu u_\mu=-1$. Therefore, the 1pdf (\ref{betazeta}) of our DTT for Bjorken flow reads
\be
f_{B}=\exp\left\{-\frac{1}{T}p^\tau+\frac{\zeta}{Tp^\tau}\left[p_x^2+p_y^2-2\frac{p_\eta^2}{\tau^2}\right] \right\}
\label{fb}
\ee
where $p^\tau=\sqrt{p_x^2+p_y^2+{p_\eta^2}/{\tau^2}}$ because of the mass shell condition and $\zeta$ is the only independent component of the tensor $\zeta^\mu_\nu=\text{diag}(0,\zeta,\zeta,-2\zeta)$ from (\ref{betazeta}).

\subsection{Dynamical equations}
Since we are interested in solving the hydrodynamical equations (\ref{DynamicEqs}), we need to compute the tensors (\ref{DTTeqs}) in terms of $\zeta$ and $T$ through the one particle distribution function $f_B$ (eq. (\ref{fb})). From the second equation of (\ref{DynamicEqs})
\be
\label{balance}
A^{\mu i}_{j;\mu}-\frac13\delta^i_jA^{\mu k}_{k;\mu}-\left[K^i_j-\frac13\delta^i_jK^k_k\right]-\left[I^i_j-\frac13\delta^i_jI^k_k\right]=0
\ee
where Latin indices are 1, 2 or 3. The $I^{\nu\rho}$ tensor (\ref{DTTeqs}) with an Anderson-Witting collision term (\ref{AW}) reads
\be
I^{\nu\rho}=-\frac{1}{\tau_R}\int \frac{Dp}{\sqrt{-g}}\;p^\nu p^\rho(f_B-f_B^{eq})
\ee
where the relaxation time $\tau_R$ is taken as $\tau_R=c/T_0(\tau)$ with $c$ a constant in order to preserve the conformal invariance (\ref{confinv}). The equilibrium 1pdf is $f_B^{eq}=\exp(-p^\tau/T_0)$. $T_0$ is defined through the Landau-Lifshitz prescription

\be 
\frac3{\pi^2}T_0^4=T^{\tau\tau}
\te
We see that the same integral defines $I^{\nu\rho}$ and the EMT, so we write
\be
I^{\nu\rho}=-\frac{T_0}{c}\left[T^{\nu\rho}-T^{\nu\rho}_{(eq)}\right]
\ee
Because the EMT is traceless and the Landau-Lifshitz prescription we have
\bea
I^k_k&=&\frac{T_0}{c}\left[T^k_{k\;(eq)}-T^k_k\right]\nn
&=&\frac{T_0}{c}\left[T^{\tau\tau}_{(eq)}-T^{\tau\tau}\right]=0
\tea
Since we only need two independent equations to compute $\zeta(\tau)$ and $T(\tau)$, we take the $\tau$ component of the EMT conservation (\ref{DynamicEqs})
\be
\partial_\tau T^{\tau\tau}+\frac1\tau\left(T^{\tau\tau}+T^\eta_\eta\right)=0
\ee
and the $\left(^\eta_\eta\right) $ component of (\ref{balance}). Observe that also
\be
A^{\tau\mu\nu}=T^{\mu\nu}
\ee
Working out the covariant derivatives explicitly we find
\bea
A^{\mu x}_{x;\mu}&=&\partial_\tau T^x_x+\frac1tT^x_x\nn
A^{\mu y}_{y;\mu}&=&\partial_\tau T^y_y+\frac1tT^y_y\nn
A^{\mu \eta}_{\eta;\mu}&=&\partial_\tau T^\eta_\eta+\frac 3\tau T^\eta_\eta
\tea
so the trace is
\be
\partial_\tau T^{\tau\tau}+\frac1\tau T^{\tau\tau}+\frac 2\tau T^\eta_\eta=\frac1\tau T^\eta_\eta
\te
and thereby
\be
A^{\mu \eta}_{\eta;\mu}-\frac13A^{\mu k}_{k;\mu}=\partial_\tau T^\eta_\eta+\frac83\frac1\tau T^\eta_\eta
\ee
The $\left(^\eta_\eta\right) $ component of $I^{\nu}_{\rho}$ is
\bea
I^\eta_\eta&=&\frac{T_0}{c}\left[T^\eta_{\eta\;(eq)}-T^\eta_\eta\right]\nn
&=&\frac{T_0}{c}\left[\frac13T^{\tau\tau}-T^\eta_\eta\right]
\tea
where $T^\eta_{\eta\;(eq)}=T^{\tau\tau}_{(eq)}/3=T^{\tau\tau}/3$ has been used. Therefore the equations of motion reads
\bea
\partial_\tau T^{\tau\tau}+\frac1\tau\left(T^{\tau\tau}+T^\eta_\eta\right)=0\nn
\partial_\tau T^\eta_\eta+\frac83\frac1\tau T^\eta_\eta -\frac23\left[K^\eta_\eta- K^x_x\right]-\frac{T_0}{c}\left[\frac13T^{\tau\tau}-T^\eta_\eta\right]=0
\label{dyn1}
\tea
We need to compute $T^{\tau\tau}$, $T^\eta_\eta$, $K^\eta_\eta$ and $K^x_x$ in terms of $T$ and $\zeta$ in order to obtain a closed dynamical system for these variables. On dimensional grounds we write

\bea
T^{\tau\tau}&=&T^4F(\zeta)\nn
T^\eta_\eta&=&T^4G(\zeta)\nn
\frac23\left[K_\eta^\eta-K_x^x\right]&=&\frac{T^4}{\tau}L(\zeta)
\label{FGL}
\tea
The functions $F$, $G$ and $L$ are derived in \ref{ctc}. Using the chain rule
\bea
\partial_\tau T^{\tau\tau}&=&4T^3F\Dot T+T^4F'\Dot\zeta\nn
\partial_\tau T^\eta_\eta&=&4T^3G\Dot T+T^4G'\Dot\zeta
\tea
where dot means $d/d\tau$ and prime $d/d\zeta$, we can rewrite the dynamic equations (\ref{dyn1}) as
\bea
\Dot \zeta &=& \frac{1}{G'-GF'/F}\left[\frac{1}{\tau}\left(\frac{G^2}{F}-\frac{5G}{3}+L\right)+\frac{T_0}{c}\left(\frac{F}{3}-G\right)\right]\nn
\Dot T&=&\frac{T}{4(G-G'F/F')}\left[\frac{1}{\tau}\left(\frac{G'(F+G)}{F'}-\frac{8G}{3}+L\right)+\frac{T_0}{c}\left(\frac{F}{3}-G\right)\right]
\label{dsys_bjor}
\tea
This is a closed dynamical system for $\zeta$ and $T$. From (\ref{FGL}) $T_0$ can be expressed as
\be
T_0=\sqrt{\pi}\;T \left(\frac{F(\zeta)}{3}\right)^{1/4}
\ee
It can be checked that the functions $G'-GF'/F$ and $G-G'F/F'$ do not vanish throughout $-1/2<\zeta<1$, so the equations are well defined in this domain.

\subsection{Exact Boltzmann equation solution}
Bjorken flow admits an exact solution of the Boltzmann equation with Anderson-Witting collision term. We follow the method of solution presented in ref. \citen{FRS13a}. 

The solution has the form
\be
\label{formalsolbjor}
f(\tau,p_x,p_y,p_\eta)=D(\tau,\tau_0)f_i(p_x,p_y,p_\eta)+\int_{\tau_0}^\tau\frac{d\tau'}{\tau_R(\tau)}D(\tau,\tau')f_{eq}(\tau',p_x,p_y,p_\eta)
\ee
where
\be
D(\tau_2,\tau_1)=\exp\left[-\int_{\tau_1}^{\tau_2}\frac{d\tau}{\tau_R(\tau)}\right]
\label{Damping}
\ee
is the so-called damping function, $f_i$ is the initial distribution function and $f_{eq}$ is the equilibrium distribution function. We assume an initial condition of the Romatschke-Strickland kind\cite{RS03}

\be
f_i(p_x,p_y,p_\eta)=\exp{\left[-\frac{1}{T_i}\sqrt{p_x^2+p_y^2+p_\eta^2/\tau_i^2}\right]}
\ee
\be
f_{eq}(\tau,p_x,p_y,p_\eta)=\exp{\left[-\frac{1}{T_0(\tau)}\sqrt{p_x^2+p_y^2+p_\eta^2/\tau^2}\right]}
\ee
where $\tau_i$ is the initial time and $T_i$ is the initial temperature. The formal solution (\ref{formalsolbjor}) is however implicit because of the $\tau$-dependence of $T_0$. To solve this, one can compute the energy density with this distribution function and use the Landau-Lifshitz condition to find an integral equation for $T_0(\tau)$:
\be
\label{EBEsol}
T_0(\tau)^4=D(\tau,\tau_0)T_i^4R\left(\frac{\tau}{\tau_0}\right)+\frac{1}{c}\int_{\tau_i}^\tau d\tau'\;D(\tau,\tau')T_0(\tau')^5R\left(\frac{\tau}{\tau'}\right)
\ee
where we used $\tau_R(\tau)=c/T_0(\tau)$ and defined
\be
R(x)=\frac12\left[\frac{1}{x^2}+\frac{\tan^{-1}[\sqrt{x^2-1}]}{\sqrt{x^2-1}}\right]
\ee
Equation (\ref{EBEsol}) can be solved by an iterative method described in ref. \citen{FRS13a}. Once $T_0(\tau)$ is computed, other $T^{\mu\nu}$ components can be obtained by taking the appropriate moment of the 1pdf (\ref{formalsolbjor}).

\subsection{Chapman-Enskog approximation}
The third order Chapman-Enskog equations for Bjorken flow are\cite{ChHPV18, J13}
\bea
\Dot \epsilon &=&-\frac{1}{\tau}\left[\frac{4}{3}\epsilon+\Pi\right]\nn
\Dot \Pi&=& -\frac{\Pi}{\tau_R}-\frac{1}{\tau}\left[\frac{16}{45}\epsilon+\frac{38}{21}\Pi-\frac{54}{49}\frac{\Pi^2}{\epsilon}\right]
\label{CE4Bjorken}
\tea
where $\Pi=\Pi^\eta_\eta$ is the only independent component of viscous EMT (\ref{viscousEMT}) and $\tau_R$ is the relaxation time. As before we take $\tau_R=c/T_0(\tau)$, with $T_0=\sqrt{\pi}(\epsilon/3)^{1/4}$. 

In this scheme, the entropy density can be written as\cite{ChHPV18}
\be
s=\frac{4T_0^3}{\pi^2}-\frac{45}{32}\frac{\Pi^2}{T_0\epsilon}+\frac{1125}{896}\frac{\Pi^3}{T_0\epsilon^2}
\label{EntropyCE}
\ee

\subsection{Numerical results}
We solved numerically the dynamical system (\ref{dsys_bjor}) and compared the results with the exact Boltzmann equation solution and the third order Chapman-Enskog approximation described above. We have used $\zeta(\tau_i)=0$ (isotropic initial configuration) and $T_i=T(\tau_i)=1$ without loss of generality. We used $\tau_i=0.25$ fm/c. For the third order Chapman-Enskog system (\ref{CE4Bjorken}) the initial conditions are $\epsilon_i=3T_i^4/\pi^2$ and $\Pi_i=0$.

The constant $c$ defined by the relaxation time $\tau_R=c/T_0$ can be rewritten as $c=5\eta/s$\cite{FRS13a}, where $\eta$ is the shear viscosity and $s$ the entropy density. We have used a specific shear viscosity $\eta/s=1/4\pi$, which saturates the Kovtun-Son-Starinets bound\cite{KSS05}.

\begin{figure}[H]
\centering
\includegraphics[scale=0.25]{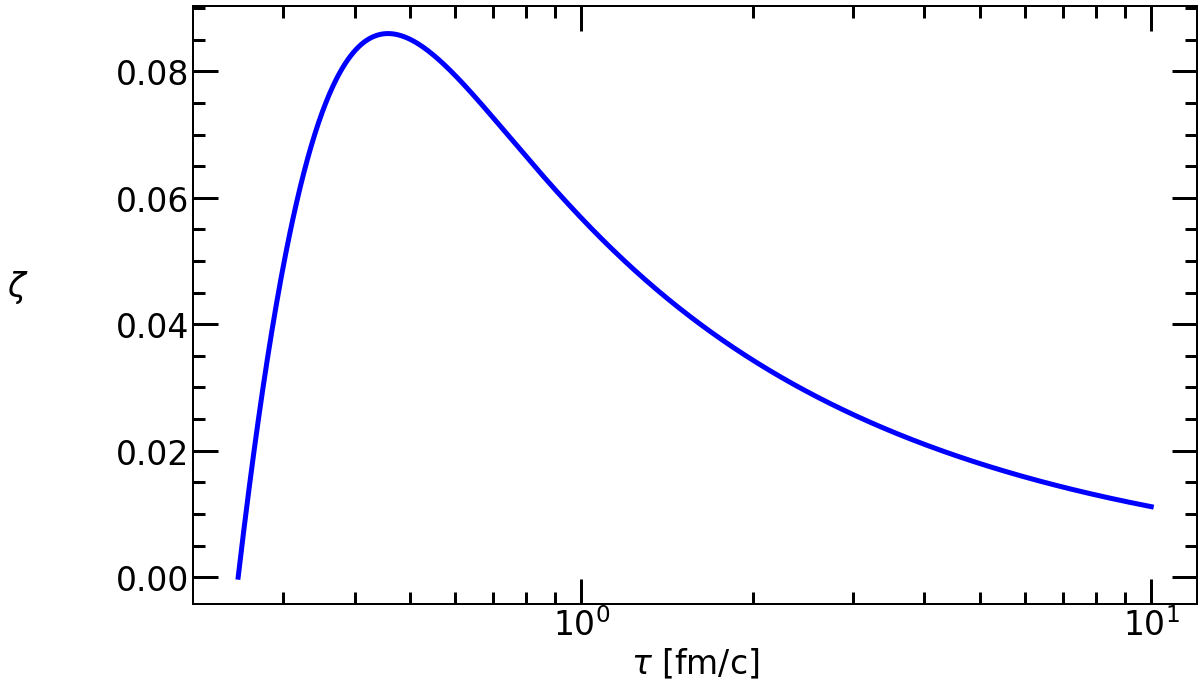}
\caption{$\zeta$ as a function of $\tau$ for Bjorken flow with specific shear viscosity $4\pi\eta/s=1$ in DTT framework.}
\label{Z_Bjorken}
\end{figure}

In Fig. \ref{Z_Bjorken} we plot $\zeta$ vs $\tau$ in semilogarithmic scale from $\tau=0.25$ to $\tau=10$ for Bjorken flow.

\begin{figure}[H]
\centering
\includegraphics[scale=0.25]{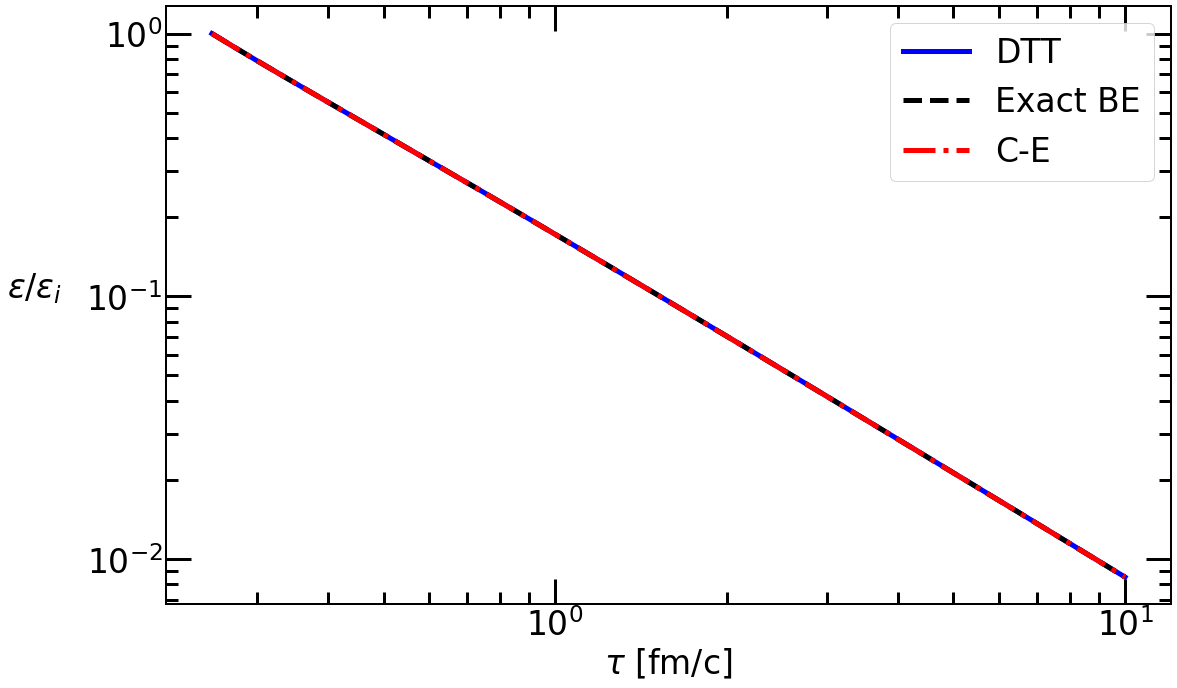}
\caption{Normalized energy density as a function of $\tau$ for Bjorken flow with specific shear viscosity $4\pi\eta/s=1$. The normalization factor is $\epsilon_i=3T_i^4/\pi^2$. Blue continuous line: DTT, black dashed line: exact Boltzmann equation, red dot-dashed line: third order Chapman-Enskog approximation.}
\label{Energia_Bjorken}
\end{figure}

In Fig. \ref{Energia_Bjorken} we plot the normalized energy density vs $\tau$ in logarithmic scale from $\tau=0.25$ to $\tau=10$ for Bjorken flow. 
Both the DTT and Chapman-Enskog curves show a strong agreement with the exact solution.

\begin{figure}[H]
\centering
\includegraphics[scale=0.25]{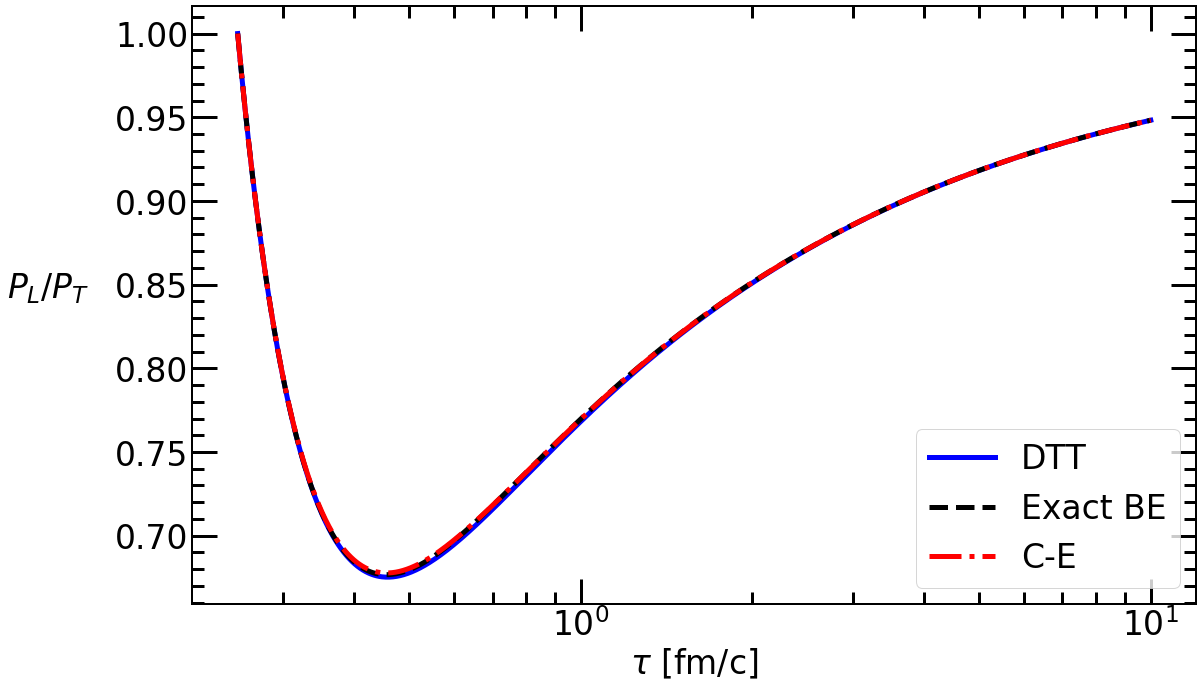}
\caption{Pressure anisotropy $P_L/P_T$ as a function of $\tau$ for Bjorken flow with specific shear viscosity $4\pi\eta/s=1$. Blue continuous line: DTT, black dashed line: exact Boltzmann equation, red dot-dashed line: third order Chapman-Enskog approximation.}
\label{PLPT_Bjorken}
\end{figure}

Fig. \ref{PLPT_Bjorken} shows the pressure anisotropy $P_L/P_T$ (where $P_T=T^x_x=T^y_x$ is the transverse pressure) as a function of $\tau$ in semilogarithmic scale from $\tau=0.25$ to $\tau=10$ for Bjorken flow. We observe again a strong agreement with the exact solution.

\begin{figure}[H]
\centering
\includegraphics[scale=0.25]{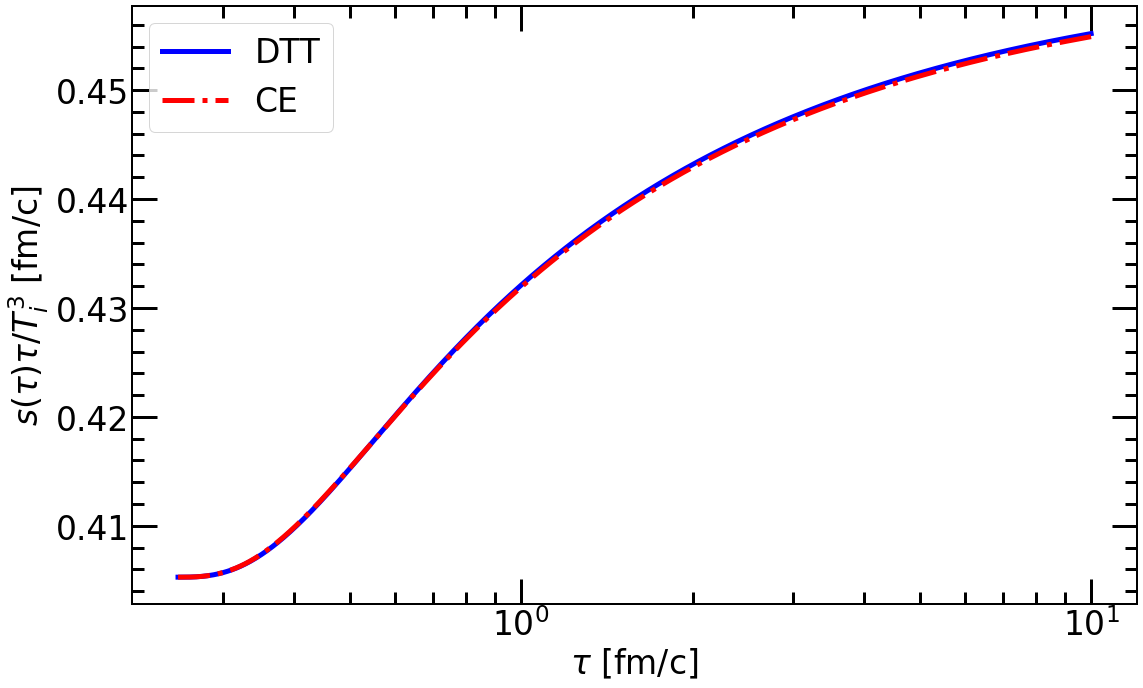}
\caption{Normalized entropy density times $\tau$, $s(\tau)\tau/T_i^3$ as a function of $\tau$ for Bjorken flow with specific shear viscosity $4\pi\eta/s=1$. Blue continuous line: DTT, red dot-dashed line: third order Chapman-Enskog approximation.}
\label{Entropia_Bjorken}
\end{figure}

In Fig. \ref{Entropia_Bjorken} we plot the normalized entropy density (see \ref{ctc}) times $\tau$, $s(\tau)\tau/T_i^3$ vs $\tau$ in semilogarithmic scale. 

\section{Gubser flow}
Gubser flow improves upon Bjorken flow in the sense that the slabs of matter are no longer homogeneous in the transverse directions. The background metric of Gubser flow is obtained through a conformal transformation of Minkowski spacetime. It can be written as
\be 
ds^2=-d\rho^2+\cosh^2\rho\left(d\theta^2+\sin^2\theta d\phi^2\right)+d\eta^2
\te
The nontrivial Christoffel symbols are
\bea
\Gamma^{\theta}_{\theta\rho}&=&\Gamma^{\phi}_{\phi\rho}=\tanh\rho\nn
\Gamma^{\theta}_{\phi\phi}&=&-\sin\theta\;\cos\theta\nn
\Gamma^{\phi}_{\phi\theta}&=&\left(\tan\theta\right)^{-1}\nn
\Gamma^{\rho}_{\theta\theta}&=&\cosh\rho\;\sinh\rho\nn
\Gamma^{\rho}_{\phi\phi}&=&\Gamma^{\rho}_{\theta\theta}\sin^2\theta
\tea
In this geometry, the 1pdf (\ref{betazeta}) of our DTT becomes
\be
f_{G}=\exp\left\{-\frac{1}{T}p^\rho+\frac{\zeta}{Tp^\rho}\left[\frac{p_\Omega^2}{\cosh^2\rho}-2p_\eta^2\right] \right\}
\ee
where $p_\Omega^2=p_\eta^2+p_\phi^2/\sin^2\theta$ and $p^\rho=\sqrt{p_\Omega^2/\cosh^2\rho+p_\eta^2}$ because of the mass shell condition. Like in the Bjorken case, $\zeta$ is the only independent component of the tensor $\zeta^\mu_\nu=\text{diag}(0,\zeta,\zeta,-2\zeta)$ from (\ref{betazeta}).

\subsection{Dynamical equations}
To obtain the dynamical equations, we need to compute $T^\eta_\eta$, $T^{\rho\rho}$, $K^\eta_\eta$ and $K^\theta_\theta$ in terms of $\zeta$ and $T$. This is done in \ref{ctcg}.

The nontrivial components of $T^{\mu\nu}$ may be written as $T^{\rho\rho}=T^4F(\zeta)$, $T^\eta_\eta=T^4G(\zeta)$, with the same $F$ and $G$ as in the Bjorken case eq. (\ref{FGL}). For the nonequilibrium tensor we find $A^{i\rho\rho}=0=A^{ijk}$, $A^{\rho\;ij}=T^{ij}$. The nontrivial covariant derivatives are ($\;\dot{} =\partial /\partial\rho$)
\bea
A^{\mu\;\theta}_{\theta;\mu}&=&\dot T^{\theta}_{\theta}+4\tanh\rho\; T^{\theta}_{\theta}\nn
A^{\mu\;\phi}_{\phi;\mu}&=&\dot T^{\phi}_{\phi}+4\tanh\rho\; T^{\phi}_{\phi}\nn
A^{\mu\;\eta}_{\eta;\mu}&=&\dot T^{\eta}_{\eta}+2\tanh\rho\; T^{\eta}_{\eta}
\tea
Taking the trace and using the EMT tracelessness condition $T^{\theta}_{\theta}+T^{\phi}_{\phi}+ T^{\eta}_{\eta}=T^{\rho\rho}$ we get
\be
A^{\mu\;k}_{k;\mu}=\dot T^{\rho\rho}+4\tanh\rho\; T^{\rho\rho}-2\tanh\rho\; T^{\eta}_{\eta}
\te  
Finally

\be
\frac{2}{3}[K^\eta_\eta-K^\theta_\theta]=T^4\tanh\rho\;L_G(\zeta)
\label{LG}
\ee
The function $L_G$ is given in eq. (\ref{LGZ}).

The DTT dynamical equations (\ref{DynamicEqs}) become
\bea
\label{dyn2}
\Dot{T}^\eta_\eta+\tanh\rho\left[\frac73T^\eta_\eta-\frac13T^{\rho\rho}\right]-\frac{2}{3}[K^\eta_\eta-K^\theta_\theta]-\frac{T_0}{c}\left[\frac{T^{\rho\rho}}{3}-T^\eta_\eta\right]&=&0\nn
\Dot{T}^{\rho\rho}+\tanh\rho\;(3T^{\rho\rho}-T^\eta_\eta)&=&0
\tea
where $T_0$ is the Landau-Lifshitz temperature, $3T_0^4/\pi^2=T^{\rho\rho}$. The system (\ref{dyn2}) becomes

\bea
\label{dsys_Gubser}
\Dot \zeta &=& \frac{1}{G'-GF'/F}\left[\tanh\rho\left(\frac23 G-\frac{G^2}{F}+\frac{F}{3}+L_G\right)+\frac{T_0}{c}\left(\frac{F}{3}-G\right)\right]
\\
\Dot T&=&\frac{T}{4(G-G'F/F')}\left[\tanh\rho\left(\frac{G'(3F-G)}{F'}-\frac{7G}{3}+\frac{F}{3}+L_G\right)+\right.\nn
&&+\left.\frac{T_0}{c}\left(\frac{F}{3}-G\right)\right]
\tea
\subsection{Exact Boltzmann equation solution}
Like in the  Bjorken case, Gubser flow has an exact Boltzmann equation formal solution in the relaxation time approximation\cite{DHMNS14b}. Computing the energy density with this solution and using the Landau-Lifshitz prescription one obtains an integral equation for $T_0(\rho)$
\be
T_0(\rho)^4=D(\rho,\rho_0)T_i^4H\left(\frac{\cosh\rho_0}{\cosh\rho}\right)+\frac{1}{c}\int_{\rho_0}^\rho d\rho'\;D(\rho,\rho')T_0(\rho')^5H\left(\frac{\cosh\rho'}{\cosh\rho}\right)
\label{ExactGubser}
\ee
where $D(\rho_2,\rho_1)$ is the damping function (\ref{Damping}), $T_i$ is the initial temperature, $c=T_0\tau_R$ and
\be
H(x)=\frac{1}{2}\left[x^2+x^4\frac{\tanh^{-1}(\sqrt{1-x^2})}{\sqrt{1-x^2}}\right]
\ee
Equation (\ref{ExactGubser}) can be solved by an iterative method\cite{DHMNS14b}.
\subsection{Chapman-Enskog approximation}
The third order Chapman-Enskog equations for Gubser flow are\cite{ChHPV18}
\bea
\partial_\rho\epsilon&=&-\left(\frac{8}{3}\epsilon-\Pi\right)\tanh\rho\nn
\partial_\rho\Pi&=&-\frac{\Pi}{\tau_R}+\tanh\rho\left(\frac{16}{45}\epsilon-\frac{46}{21}\Pi-\frac{54}{49}\frac{\Pi^2}{\epsilon}\right)
\tea
where $\epsilon=T^{\rho\rho}$ is the energy density, $\Pi=\Pi^\eta_\eta$ is the only independent component of the viscous EMT (\ref{viscousEMT}) and the relaxation time $\tau_R$ is taken as $\tau_R(\rho)=c/T_0(\rho)$, with $T_0=\sqrt{\pi}(\epsilon/3)^{1/4}$.
The entropy density has the same expression as Bjorken (\ref{EntropyCE}), but inserting the Gubser values for $\Pi$, $\epsilon$ and $T_0$ \cite{ChHPV18}.

\subsection{Numerical results}
We solved numerically the dynamical system (\ref{dsys_Gubser}) and we compared the solution with the exact Boltzmann equation solution and the third order Chapman-Enskog approximation described above. We have used $\zeta(\rho_i)=0$ (isotropic initial configuration) and $T(\rho_i)=0.002$ with $\rho_i=-10$ as in ref. \citen{MMcNH17}. For the third order Chapman-Enskog system (\ref{CE4Bjorken}) the initial conditions are $\epsilon_i=3T_i^4/\pi^2$ and $\Pi_i=0$. We also used a specific shear viscosity $4\pi\eta/s=1$.

\begin{figure}[H]
\centering
\includegraphics[scale=0.25]{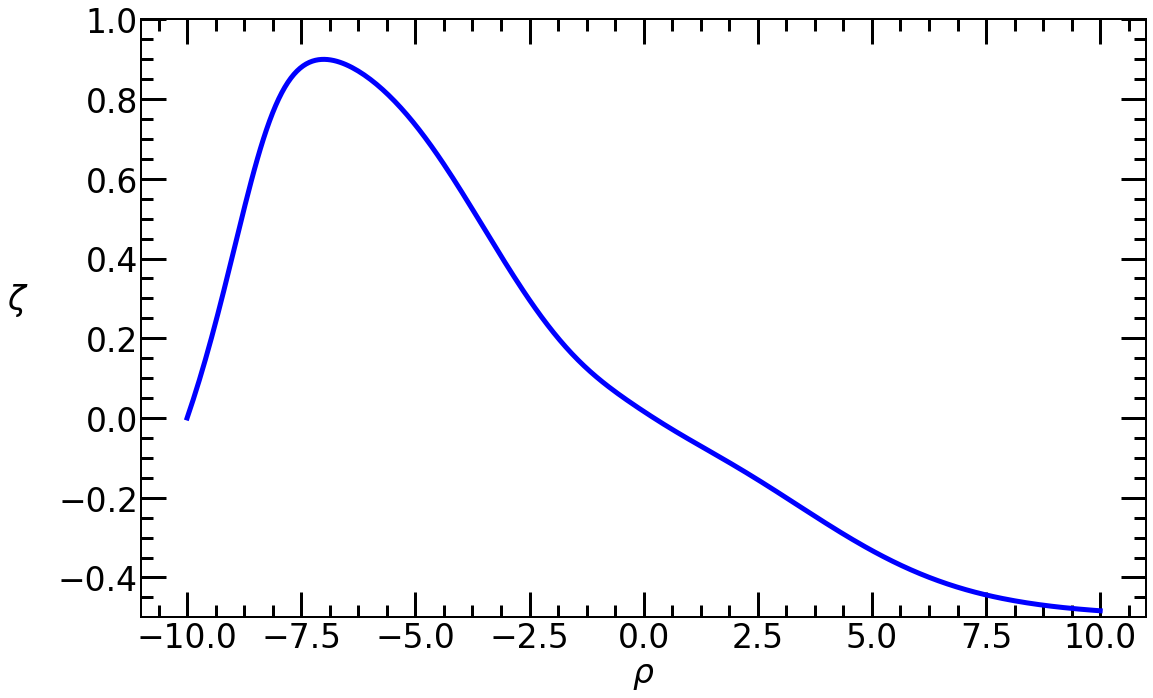}
\caption{$\zeta$ as a function of $\rho$ for Gubser flow in the DTT framework with specific shear viscosity $4\pi\eta/s=1$.}
\label{Z_Gubser}
\end{figure}

In Fig. \ref{Z_Gubser} we plot $\zeta$ vs $\rho$ in natural scale from $\rho=-10$ to $\rho=10$ for Gubser flow. Note that the $\zeta(\rho)$ curve is qualitatively similar to the anisotropy parameter $\xi(\rho)$ from anisotropic hydrodinamics defined in ref. \citen{MMcNH17}.

\begin{figure}[H]
\centering
\includegraphics[scale=0.25]{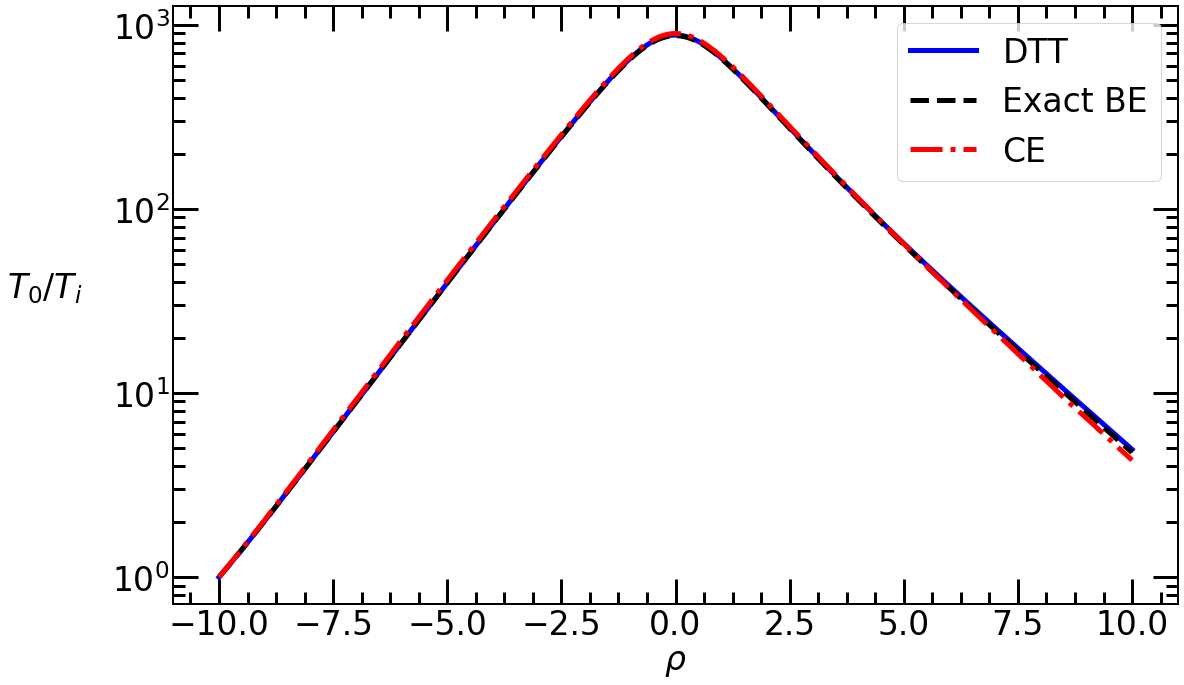}
\caption{Normalized Landau temperature $T_0/T_i$ as a function of $\rho$ for Gubser flow with specific shear viscosity $4\pi\eta/s=1$. Blue continuous line: DTT, black dashed line: exact Boltzmann equation, red dot-dashed line: third order Chapman-Enskog approximation.}
\label{Temperatura_Gubser}
\end{figure}

In Fig. \ref{Temperatura_Gubser} we plot the normalized Landau Temperature $T_0/T_i$ vs $\rho$ in semilogarithmic scale from $\rho=-10$ to $\rho=10$ for Gubser flow. All three theories agree closely but for large values of $\rho$ DTT is closer to the exact solution.

\begin{figure}[H]
\centering
\includegraphics[scale=0.25]{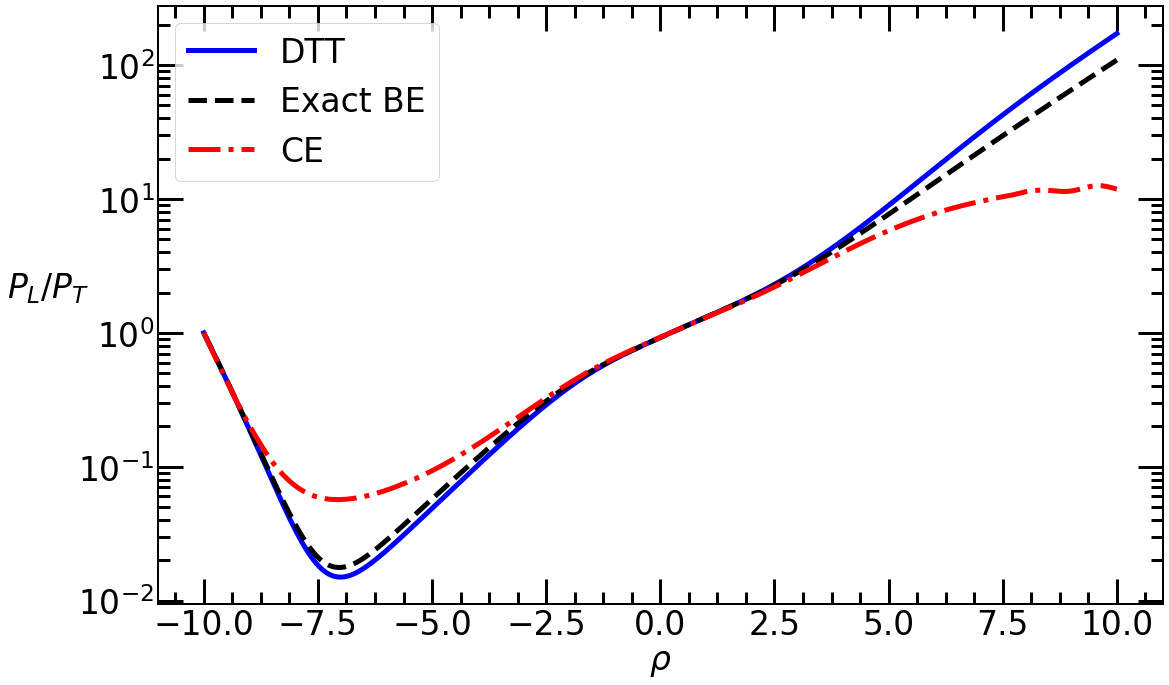}
\caption{Pressure anisotropy $P_L/P_T$ as a function of $\rho$ for Gubser flow with specific shear viscosity $4\pi\eta/s=1$. Blue continuous line: DTT, black dashed line: exact Boltzmann equation, red dot-dashed line: third order Chapman-Enskog approximation.}
\label{PLPT_Gubser}
\end{figure}

Fig. \ref{PLPT_Gubser} shows the pressure anisotropy $P_L/P_T$ (where $P_T=T^\theta_\theta=T^\phi_\phi$ is the transverse pressure) as a function of $\rho$ in semilogarithmic scale from $\rho=-10$ to $\rho=10$. The DTT curve significantly improves upon the Chapman-Enskog approximation.

\begin{figure}[H]
\centering
\includegraphics[scale=0.25]{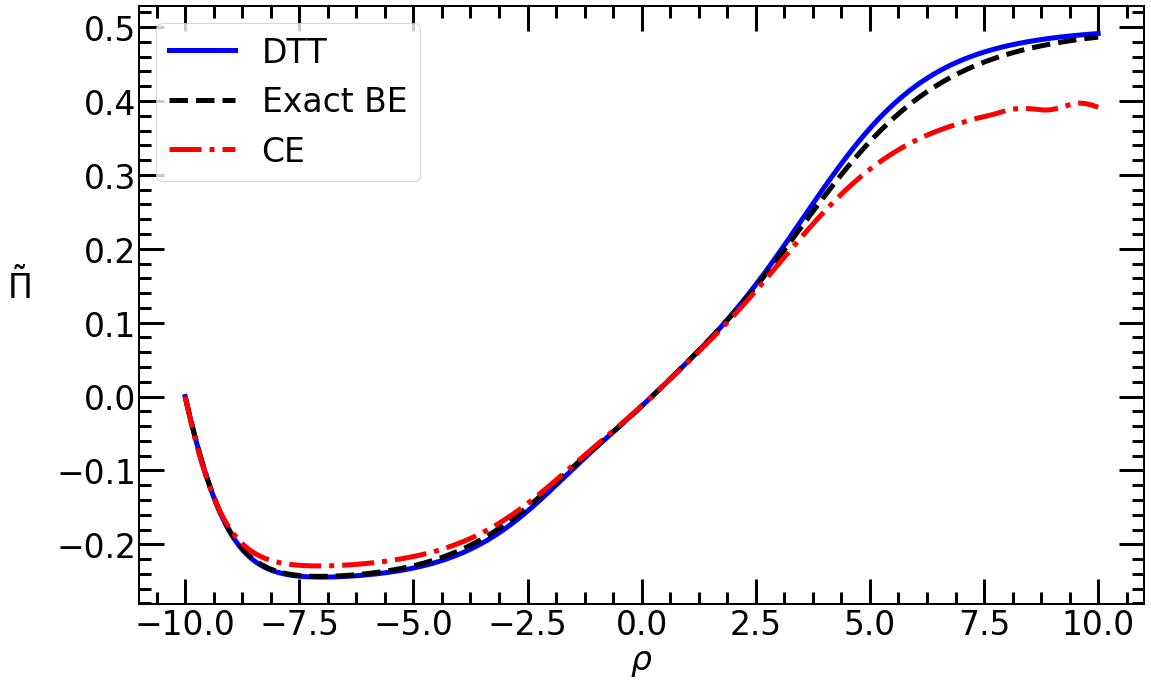}
\caption{Normalized shear stress (defined as $\Tilde\Pi=3\Pi^\eta_\eta/(4\epsilon)$) vs $\rho$ for Gubser flow with specific shear viscosity $4\pi\eta/s=1$. Blue continuous line: DTT, black dashed line: exact Boltzmann equation, red dot-dashed line: third order Chapman-Enskog approximation.}
\label{Pi_Gubser}
\end{figure}

In Fig. \ref{Pi_Gubser} we show the normalized shear stress defined as $\Tilde\Pi=3\Pi^\eta_\eta/(4\epsilon)$ vs $\rho$. We observe a higher agreement between the DTT and the exact one than between the latter and the Chapman-Enskog approximation. 

\begin{figure}[H]
\centering
\includegraphics[scale=0.25]{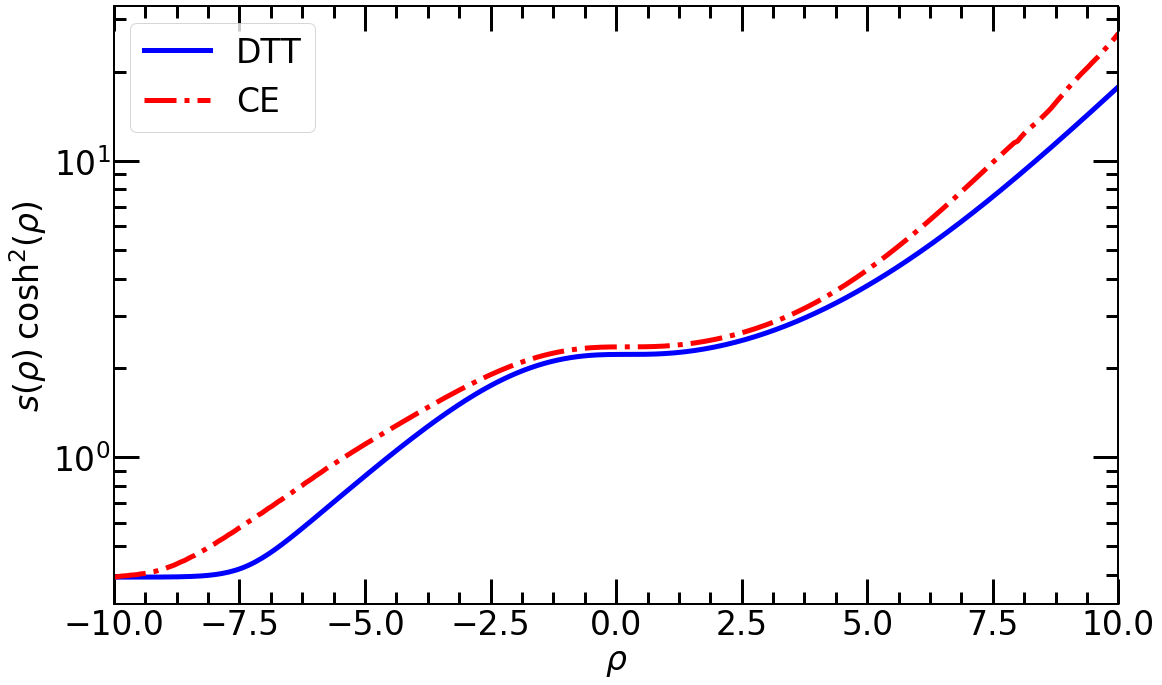}
\caption{Normalized entropy density $s(\rho)$ times $\cosh^2\rho$ as a function of $\rho$ for Gubser flow with specific shear viscosity $4\pi\eta/s=1$. Blue continuous line: DTT, red dot-dashed line: third order Chapman-Enskog approximation.}
\label{Entropia_Gubser}
\end{figure}

In Fig. \ref{Entropia_Gubser} we plot the entropy density $s(\rho)$  (see \ref{ctcg}) times $\cosh^2\rho$ vs $\rho$ in semilogarithmic scale. A good agreement between the DTT and Chapman-Enskog curves is observed.

\section{Conclusions}
In this paper we have shown that the requirement of thermodynamic consistency to all orders in deviations from equilibrium practically singles out a DTT framework\cite{L72,LMR86,GL90,GL91,NR95,RN97,BR99,M99,C98,CT01,PRC09,PRC10a,PRC10c,PRC10b,PRC12,MGC17,LRR18} as the proper relativistic replacement for the Navier-Stokes equations, and then that the EPVM\cite{PRC10a,PRC13a,Cal13a}  may be fruitfully used to single out a particular DTT. The resulting theory performs well in the Bjorken and Gubser cases, being simpler than several competitive alternatives. Moreover it is framed in a fully covariant way and it may be easily generalized to more general backgrounds and to quantum statistics\cite{AC17}.

The formal device of introducing \emph{two} vector fields $u^{\mu}$ and $U^{\mu}$, where the former is the hydrodynamic degree of freedom while the latter is regarded as an external parameter, to be identified \emph{after} the equations of motion are derived, has been used many times in the literature, most notably in the quantization of non abelian gauge theories\cite{Abbott81,Hart83}.

It is clear that many challenges remain ahead, such as to generalize the theory to realistic collision terms\cite{BMH16}, to go beyond conformal invariance\cite{BHM15}, and finally to use the formalism in actual problems\cite{MGC17,ANRS17}. We expect to be able to report on progress in these directions in the near future.

\section*{Acknowledgments}
Cristi\'an Vega and Fernando Paz collaborated in early stages of this project. The work of LC and EC was supported in part by
CONICET, ANPCyT and University of Buenos Aires. LC is supported by a fellowship from CONICET.

\appendix

\section{Relativistic phase space}
\label{rps}

In this Appendix we shall expand on some properties of the phase space of a relativistic particle which are relevant to our discussion. Phase space is $M\times \mathbf{R}^4$, where $M$ is the space time manifold and $ \mathbf{R}^4$ its tangent space. A tensor field $X^{\mu_1\ldots\mu_n}\left(x,p_{\nu}\right)$ in phase space transforms under a coordinate change $x\to x'$ as

\be
X^{\mu_1\ldots\mu_n}\left(x,p_{\nu}\right)\to X'^{\mu'_1\ldots\mu'_n}\left(x',p'_{\nu'}\right)=\prod_{j=1}^n\frac{\partial x'^{\mu'_j}}{\partial x^{\mu_j}}X^{\mu_1\ldots\mu_n}\left(x,\frac{\partial x'^{\nu'}}{\partial x^{\nu}}p'_{\nu'}\right)
\te 
For example, if $\beta^{\mu}\left(x\right)$ is an spacetime vector, then $\beta^{\mu}p_{\mu}$ is a phase space scalar.

The covariant derivative of a scalar $\mathcal{R}\left(x,p\right)$ is defined by the operator

\be
\nabla_{\mu}\mathcal{R}=\frac{\partial \mathcal{R}}{\partial x^{\mu}}+\Gamma^{\rho}_{\mu\sigma}p_{\rho}\frac{\partial \mathcal{R}}{\partial p_{\sigma}}
\label{nabla}
\te
where $\Gamma$ is the connection. This covariant derivative defines a vector field. The covariant derivative of higher tensor fields is defined by requesting that the Leibnitz rule holds, and that it reduces to the ordinary covariant derivative for momentum-independent tensors. 

Momentum space is endowed with the invariant measure (later on we shall further multiply it by $2/\left(2\pi\right)^3$)

\be
\frac{d^4p_{\nu}}{\sqrt{-g}}
\te 
If $X^{\mu_1\ldots\mu_n}\left(x,p_{\nu}\right)$ is a phase space tensor, then

\be
\mathcal{X}^{\mu_1\ldots\mu_n}\left(x\right)=\int\;\frac{d^4p_{\nu}}{\sqrt{-g}}\;X^{\mu_1\ldots\mu_n}\left(x,p_{\nu}\right)
\te 
is a spacetime tensor.

A one particle distribution function is a non-negative scalar concentrated on a future oriented mass shell. This means it has the form

\be
F\left(x,p\right)=\delta\left(p^2+m^2\right)\theta\left(p^0\right)f\left(x,p\right)
\te
The mass shell projector $\delta\left(p^2+m^2\right)\theta\left(p^0\right)$ obeys 

\be
\nabla_{\mu}\delta\left(p^2+m^2\right)\theta\left(p^0\right)=0
\label{msc}
\te 
for every positive $m^2$, and so also in the $m^2\to 0$ limit, which we shall assume from now on. For this reason it is best to extract it and to define the measure

\be 
Dp=2\frac{d^4p_{\nu}}{\left( 2\pi\right)^3}\delta\left( p^2\right) \theta\left( p^0\right) =\frac{d^3p_{j}}{\left( 2\pi\right)^3p^0}
\label{invariant}
\te 
Now consider a tensor of the form

\be
A^{\mu}_{\mu_1\ldots\mu_n}\left(x\right)=\int\;\frac{d^4p_{\nu}}{\sqrt{-g}}\;p^{\mu}p_{\mu_1}\ldots p_{\mu_n}\mathcal{A}\left(x,p_{\nu}\right)
\te 
where $\mathcal{A}$ is a scalar. Then

\be
\nabla_{\mu}A^{\mu}_{\mu_1\ldots\mu_n}\left(x\right)=\frac1{{\sqrt{-g}}}\partial_{\mu}{\sqrt{-g}}A^{\mu}_{\mu_1\ldots\mu_n}-\sum_j\Gamma^{\tau}_{\mu\mu_j}A^{\mu}_{\tau\mu_1\ldots\left(\mu_j\right)\ldots\mu_n}
\te
(meaning that $\mu_j$ is omitted)

\be
=\int\;\frac{d^4p_{\nu}}{\sqrt{-g}}\;\left\{p_{\mu_1}\ldots p_{\mu_n}\left[g^{\mu\lambda}_{,\mu}p_{\lambda}\mathcal{A}+p^{\mu}\mathcal{A}_{,\mu}\right]-p^{\mu}p_{\tau}\sum_j\Gamma^{\tau}_{\mu\mu_j}p_{\mu_1}\ldots\left(p_{\mu_j}\right)\ldots p_{\mu_n}\mathcal{A}\right\}
\te
but 

\be
\sum_j\Gamma^{\tau}_{\mu\mu_j}p_{\mu_1}\ldots\left(p_{\mu_j}\right)\ldots p_{\mu_n}=\Gamma^{\tau}_{\mu\sigma}\frac{\partial}{\partial p_{\sigma}}\left[p_{\mu_1}\ldots p_{\mu_n}\right]
\te
Integrating by parts, 

\be
=\int\;\frac{d^4p_{\nu}}{\sqrt{-g}}\;p_{\mu_1}\ldots p_{\mu_n}\left\{\left[g^{\mu\lambda}_{,\mu}p_{\lambda}+\Gamma^{\tau}_{\mu\sigma}g^{\mu\sigma}p_{\tau}+\Gamma^{\tau}_{\mu\tau}p^{\mu}\right]\mathcal{A}+p^{\mu}\nabla_{\mu}\mathcal{A}\right\}
\te
The square brackets in the first term vanish and finally

\be
\nabla_{\mu}A^{\mu}_{\mu_1\ldots\mu_n}\left(x\right)=\int\;\frac{d^4p_{\nu}}{\sqrt{-g}}\;p_{\mu_1}\ldots p_{\mu_n}p^{\mu}\nabla_{\mu}\mathcal{A}
\label{lema1}
\te 
If moreover

\be 
\mathcal{A}=\delta\left( p^2+m^2\right) \theta\left( p^0\right) \mathcal{R}
\te 
from eq. (\ref{msc}) we get the more definite result

\be
\nabla_{\mu}A^{\mu}_{\mu_1\ldots\mu_n}\left(x\right)=\int\;\frac{Dp}{\sqrt{-g}}\;p_{\mu_1}\ldots p_{\mu_n}p^{\mu}\nabla_{\mu}\mathcal{R}
\label{lema}
\te 
We use the identity (\ref{lema}) with $\mathcal{R}=R_nf$ to get

\be 
\int\;\frac{Dp}{\sqrt{-g}}\;p_{\mu_1}\ldots p_{\mu_n}R_n\;p^{\mu}\nabla_{\mu}f=\nabla_{\mu}A^{\mu}_{\mu_1\ldots\mu_n}-K_{\mu_1\ldots\mu_n}
\te 
with $A^{\mu}_{\mu_1\ldots\mu_n}$ and $K_{\mu_1\ldots\mu_n}$ as in eq. (\ref{hydroconrels}).

This equation allows us to compute moments of the transport equation. It may be extended by linearity to arbitrary tensors.

Next consider the ansatz eq. (\ref{fDTT}) for the 1pdf.  By taking moments of the transport equation we get eqs. (\ref{hydrocons}) and (\ref{hydroeqs}). From eq. (\ref{lema}), the generating function eq. (\ref{gen2}) obeys 

\be
\Phi^{\mu}_{;\mu}=\sum\left[\zeta^{\mu_1\ldots\mu_n}_{;\mu}A^{\mu}_{\mu_1\ldots\mu_n}+\zeta^{\mu_1\ldots\mu_n}K_{\mu_1\ldots\mu_n}\right]
\te
and so 

\bea
S^{\mu}_{;\mu}&=&\sum\left\{\zeta^{\mu_1\ldots\mu_n}_{;\mu}A^{\mu}_{\mu_1\ldots\mu_n}+\zeta^{\mu_1\ldots\mu_n}K_{\mu_1\ldots\mu_n}-\zeta^{\mu_1\ldots\mu_n}_{;\mu}A^{\mu}_{\mu_1\ldots\mu_n}\right. \nn
&-&\left. \zeta^{\mu_1\ldots\mu_n}\left[I_{\mu_1\ldots\mu_n}+K_{\mu_1\ldots\mu_n}\right]\right\}\nn
&=&-\sum\zeta^{\mu_1\ldots\mu_n}I_{\mu_1\ldots\mu_n}
\tea

\section{Tensor components for Bjorken flow}
\label{ctc}

\begin{figure}[H]
\centering
\includegraphics[scale=0.25]{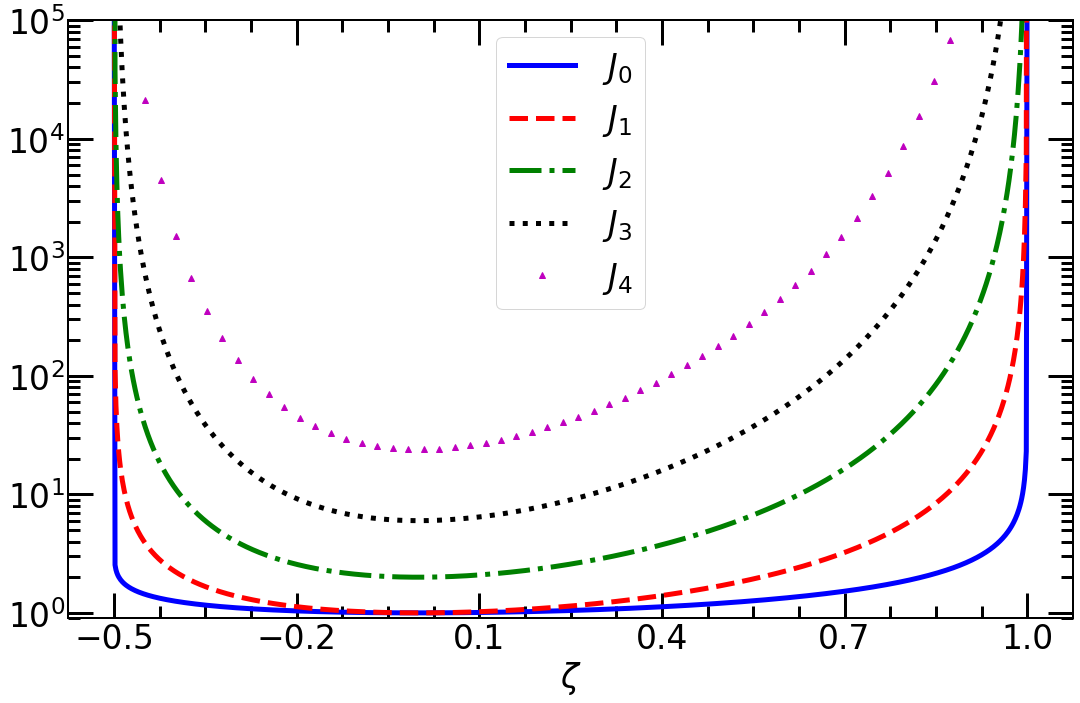}
\caption{$J_k$ functions for $k=$0, 1, 2, 3 and 4 vs $\zeta$.}
\label{Jsgraph}
\end{figure}

In this Appendix we will detail the calculation of the relevant tensors in the Bjorken flow. 
The energy density $\epsilon=T^{\tau\tau}$ is
\bea
T^{\tau\tau}&=&\frac{1}{\tau}\int \frac{d^3p}{(2\pi)^3}\;p^\tau f_B\nn
&=&\frac{T^4}{2\pi^2}\int_0^\infty dq\;q^3\int_0^1 dx\; e^{-q+\zeta q(1-3x^2)}
\tea
The longitudinal pressure $P_L=T^\eta_\eta$ is
\bea
T^{\eta}_\eta&=&\frac{1}{\tau}\int \frac{d^3p}{(2\pi)^3p^\tau}\;p^\eta p_\eta f_B\nn
&=&\frac{T^4}{2\pi^2}\int_0^\infty dq\;q^3\int_0^1 dx\; x^2\; e^{-q+\zeta q(1-3x^2)}
\label{T_eta}
\tea
The tensor component $K^\eta_\eta$ is
\bea
\label{K_eta}
K^\eta_\eta&=&\frac{T^4}{\tau}\int\frac{d^3q}{(2\pi)^3}\frac{q_z^4}{q^3}f_B\nn
&=&\frac{T^4}{\tau 2\pi^2}\int_0^\infty dq\;q^3\int_0^1 dx\;x^4\;e^{-q+\zeta q(1-3x^2)}
\tea
and $K_x^x$ is:
\bea
\label{K_xx}
K^x_x&=&\frac{T^4}{\tau}\int\frac{d^3q}{(2\pi)^3}\frac{q_z^2q_x^2}{q^3}f\nn
&=&\frac{T^4}{\tau 4\pi^2}\int_0^\infty dq\;q^3\int_0^1 dx\;(1-x^2)x^2\;e^{-q+\zeta q(1-3x^2)}
\tea
So, the problem reduces to compute the integrals
\be
J_k(\zeta)=\int_0^{\infty}dq\;q^k\int_0^{1}dx\;e^{-q+\zeta q\left(1-3x^2\right)}
\te
for $k=1$, $k=2$, $k=3$ and $k=4$ as we shall see. It is enough to compute
\bea
J_0(\zeta)&=&\int_0^{\infty}dq\;\int_0^{1}dx\;e^{-q+\zeta q\left(1-3x^2\right)}\nn
&=&\begin{cases}
\frac{1}{\sqrt{3\zeta(\zeta-1)}}\tanh^{-1}\left[\sqrt{\frac{3\zeta}{\zeta-1}}\right] & -1/2<\zeta<0\\
1 & \zeta=0\\
\frac{1}{\sqrt{3\zeta(1-\zeta)}}\tan^{-1}\left[\sqrt{\frac{3\zeta}{1-\zeta}}\right] & 0<\zeta<1
\end{cases}
\tea
and use the following recurrence relation to compute the higher $J_k$ functions
\be
J_{k+1}=\left(k+1\right)J_k+\zeta\frac{\partial}{\partial\zeta}J_k
\label{recurrence}
\te
We get
\bea
J_1&=&\frac12J_0\frac1{1-\zeta}+\frac16\frac1{\zeta+\frac12}+\frac16\frac1{1-\zeta}
\tea
\bea
J_2&=&J_1\left[ 2+\frac12\frac1{1-\zeta}\right] 
+J_0\left[ -\frac1{1-\zeta}+\frac12\frac1{\left( 1-\zeta\right) ^2}\right] \nn
&-&\frac16\frac1{\zeta+\frac12}+\frac1{12}\frac1{\left( \zeta+\frac12\right) ^2}-\frac16\frac1{1-\zeta}+\frac16\frac1{\left( 1-\zeta\right) ^2}
\tea
\bea
J_3&=&J_2\left[ 5+\frac12\frac1{1-\zeta}\right] \nn
&+&J_1\left[-4-\frac52\frac1{1-\zeta}+\frac1{\left( 1-\zeta\right) ^2} \right] \nn
&+&J_0\left[ 2\frac1{1-\zeta}-\frac52\frac1{\left( 1-\zeta\right) ^2}+\frac1{\left( 1-\zeta\right) ^3}
\right] \nn
&+&\frac16\frac1{\zeta+\frac12}-\frac1{4}\frac1{\left( \zeta+\frac12\right) ^2}+\frac1{12}\frac1{\left( \zeta+\frac12\right) ^3}\nn 
&+&\frac16\frac1{1-\zeta}-\frac12\frac1{\left( 1-\zeta\right) ^2}+\frac13\frac1{\left( 1-\zeta\right) ^3}
\tea
\bea
J_4&=&J_3\left[ 9+\frac12\frac1{1-\zeta}\right] \nn
&+&J_2\left[-19-\frac92\frac1{1-\zeta}+\frac32\frac1{\left( 1-\zeta\right) ^2} \right] \nn
&+&J_1\left[ 8+\frac{19}2\frac1{1-\zeta}-9\frac1{\left( 1-\zeta\right) ^2}+3\frac1{\left( 1-\zeta\right) ^3}\right] \nn
&+&J_0\left[ -4\frac1{1-\zeta}+\frac{19}2\frac1{\left( 1-\zeta\right) ^2}-9\frac1{\left( 1-\zeta\right) ^3}+3\frac1{\left( 1-\zeta\right) ^4}\right] \nn
&-&\frac16\frac1{\zeta+\frac12}+\frac7{12}\frac1{\left( \zeta+\frac12\right) ^2}-\frac1{2}\frac1{\left( \zeta+\frac12\right) ^3}+\frac18\frac1{\left( \zeta+\frac12\right)^4}\nn 
&-&\frac16\frac1{1-\zeta}+\frac76\frac1{\left( 1-\zeta\right) ^2}-2\frac1{\left( 1-\zeta\right) ^3}+\frac1{\left( 1-\zeta\right) ^4}
\tea
In Fig. \ref{Jsgraph} we plot the $J_k$ functions for $k=$0, 1, 2, 3 and 4 vs $\zeta$ in semilogarithmic scale. All of these functions are positive and have vertical asymptotes at $\zeta=-1/2$ and $\zeta=1$.

\begin{figure}[H]
\centering
\includegraphics[scale=0.25]{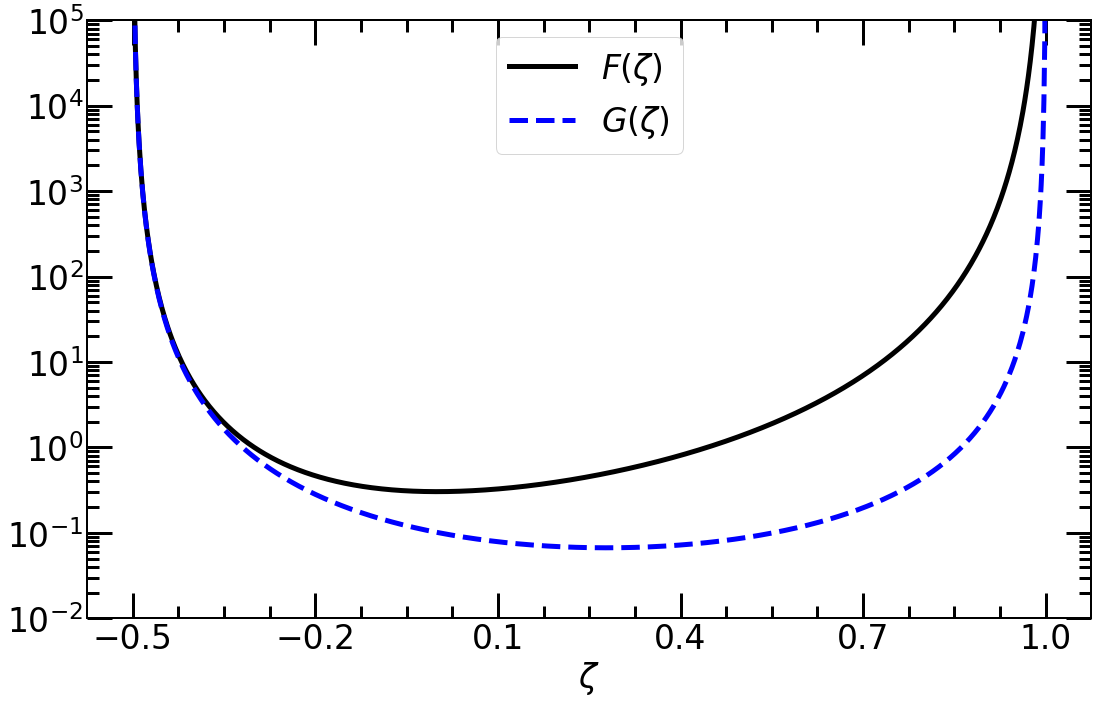}
\caption{$F$ and $G$ functions (see eqs. (\ref{FJ}) and (\ref{GJ})) vs $\zeta$.}
\label{FyG}
\end{figure}

It is immediate that
\be
T^{\tau\tau}=\frac{T^4}{2\pi^2}J_3
\ee
Therefore $F$ in eq. (\ref{FGL}) is 

\be 
F=\frac1{2\pi^2}J_3
\label{FJ}
\te
Using $x^2=\frac13-\frac13(1-3x^2)$ in (\ref{T_eta}) and integrating by parts we have
\be
T^\eta_\eta=\frac{T^4}{6\pi^2}\left[\frac{3}{\zeta}J_2+\left(1-\frac{1}{\zeta}\right)J_3\right]
\te
therefore $G$ in eq. (\ref{FGL})
\be 
G=\frac{1}{6\pi^2}\left[\frac{3}{\zeta}J_2+\left(1-\frac{1}{\zeta}\right)J_3\right]
\label{GJ}
\te
In Fig. \ref{FyG} we plot $F$ and $G$ functions vs $\zeta$ in semilogarithmic scale from $\zeta=-0.5$ to $\zeta=1$. These functions inherit the asymptotic properties of $J_k$ functions.
Note that the derivatives of $F$ and $G$ can also be expressed in terms of $J_k$ functions through relation (\ref{recurrence}) as
\bea
F'&=&\frac{1}{2\pi^2}\frac{1}{\zeta}\left[J_4-4J_3\right]\nn
G'&=&\frac{1}{6\pi^2}\frac{1}{\zeta}\left[-\frac{12}{\zeta}J_2+\left(\frac8\zeta-4\right)J_3+\left(1-\frac{1}{\zeta}\right)J_4\right]
\label{FpGp}
\tea

\begin{figure}[H]
\centering
\includegraphics[scale=0.25]{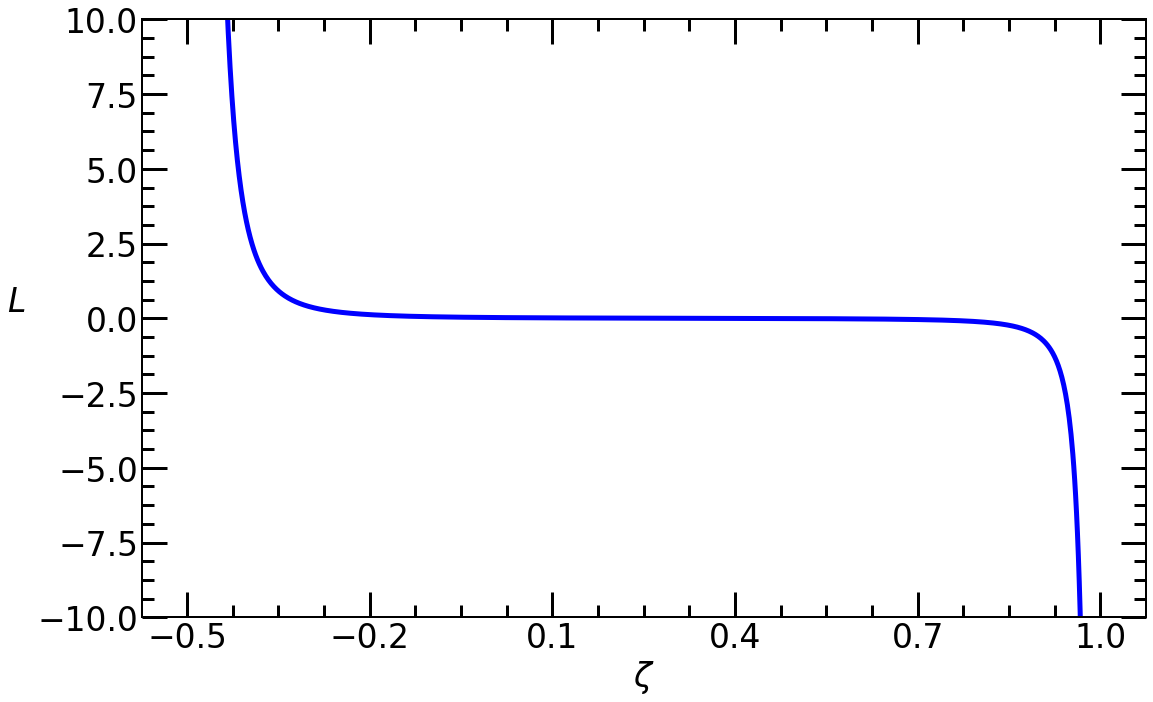}
\caption{$L$ function (see eq. (\ref{LZB})) vs $\zeta$.}
\label{LBjor}
\end{figure}

Using $x^4=\frac19-\frac29(1-3x^2)+\frac19(1-3x^2)^2$ in (\ref{K_eta}) and $(1-x^2)x^2=\frac29-\frac19(1-3x^2)-\frac19(1-3x^2)^2$ in (\ref{K_xx}) and integrating by parts two times, we obtain
\bea
K^\eta_\eta&=&\frac{T^4}{18 \pi^2\tau}\left[\frac{6}{\zeta^2}J_1+6\left(\frac{1}{\zeta}-\frac{1}{\zeta^2}\right)J_2+\left(\frac{1}{\zeta^2}-\frac{2}{\zeta}+1\right)J_3\right]\nn
K^x_x&=&\frac{T^4}{36\pi^2\tau}\left[-\frac{6}{\zeta^2}J_1+\left(\frac{3}{\zeta}+\frac{6}{\zeta^2}\right)J_2-\left(\frac{1}{\zeta}+\frac{1}{\zeta^2}+2\right)J_3\right]
\tea
and thereby $L$ from eq. (\ref{FGL}) reads
\be
L(\zeta)=\frac{1}{18\pi^2}\left[\frac{6}{\zeta^2}J_1+\left(\frac{3}{\zeta}-\frac{6}{\zeta^2}\right)J_2+\left(\frac{1}{\zeta^2}-\frac{1}{\zeta}\right)J_3\right]
\label{LZB}
\ee
This function is plotted in Fig. \ref{LBjor} vs $\zeta$ in natural scale from $\zeta=-1/2$ to $\zeta=1$. $L(\zeta)$ virtually vanishes in its domain although it rapidly tends to $\infty$ as $\zeta\to-1/2$ and to $-\infty$ as $\zeta\to1$.

The $\Phi^\mu$ function defined in (\ref{gen2}) can also be expressed in terms of $J_k$ functions. We get
\bea
\Phi^\tau&=&\frac{1}{\tau}\int \frac{d^3p}{(2\pi)^3}\; f_B\nn
&=&\frac{T^3}{2\pi^2}\int_0^\infty dq\;q^2\int_0^1 dx\; e^{-q+\zeta q(1-3x^2)}\nn
&=&\frac{T^3}{2\pi^2}J_2
\label{phiJ2}
\tea
So, the entropy density $s$ defined form entropy current $S^\mu=su^\mu$ results (see eq. (\ref{Massieu}))
\bea
\label{entropyB}
s&=&\frac{T^3}{2\pi^2}J_2+\frac{1}{T}T^{\tau\tau}-\frac{\zeta}{T}(T^x_x+T^y_y-2T^\eta_\eta)\nn
&=&\frac{T^3}{2\pi^2}J_2+\frac{1}{T}\left[T^{\tau\tau}(1-\zeta)+3\zeta T^\eta_\eta\right]\nn
&=&\frac{2T^3}{\pi^2}J_2
\tea

\section{Tensor components in Gubser flow}
\label{ctcg}

In this Appendix we expand on the calculation of the relevant tensors in Gubser flow. 

To begin with, observe that the $\zeta$ dependence of tensor components in Gubser flow may be written in terms of the same $J_k$ functions we have introduced 
in \ref{ctc}. The energy density $\epsilon=T^{\rho\rho}$ is
\bea
T^{\rho\rho}&=&\frac{1}{\cosh^2\rho\sin\theta}\int \frac{d^3p}{(2\pi)^3} p^\rho f_G\nn
&=&\frac{T^4}{2\pi^2}J_3
\tea
The longitudinal pressure $P_L=T^\eta_\eta$ is
\bea
T^\eta_\eta&=&\frac{1}{\cosh^2\rho\sin\theta}\int \frac{d^3p}{(2\pi)^3} \frac{p^\eta p_\eta}{p^\rho} f_G\nn
&=&\frac{T^4}{6\pi^2}\left[\frac{3}{\zeta}J_2+\left(1-\frac{1}{\zeta}\right)J_3\right]
\tea
The tensor component $K^\eta_\eta$ is
\bea
K^\eta_\eta&=&T^4\tanh{\rho}\int\frac{d^3q}{(2\pi)^2}\frac{q_\theta^2+q_\phi^2}{q^3}q_\eta^2f_G\nn
&=&\frac{T^4\tanh\rho}{2\pi^2}\int_0^\infty dq\;q^3\int_0^1dx\;(1-x^2)x^2\;e^{-q+q\zeta(1-3x^2)}\nn
&=&\frac{T^4\tanh\rho}{18\pi^2}\left[-\frac{6}{\zeta^2}J_1+\left(\frac{3}{\zeta}+\frac{6}{\zeta^2}\right)J_2+\left(2-\frac{1}{\zeta}-\frac{1}{\zeta^2}\right)J_3\right]
\tea
And $K^\theta_\theta$ is
\bea
K^\theta_\theta&=&T^4\tanh{\rho}\int\frac{d^3q}{(2\pi)^2}\frac{q_\theta^2+q_\phi^2}{q^3}q_\theta^2f_G\nn
&=&\frac{T^4\tanh\rho}{(2\pi)^2}\int_0^\infty dq\;q^3\int_0^1dx\;(1-x^2)^2e^{-q+q\zeta(1-3x^2)}\nn
&=&\frac{T^4\tanh\rho}{18\pi^2}\left[\frac{3}{\zeta^2}J_1-\left(\frac6\zeta+\frac{3}{\zeta^2}\right)J_2+\left(2+\frac2\zeta+\frac{1}{2\zeta^2}\right)J_3\right]
\tea
Thereby $F$ and $G$ are given by eqs. (\ref{FJ}) and (\ref{GJ}), while

\be
L_G(\zeta)=\frac{1}{9\pi^2}\left[-\frac{3}{\zeta^2}J_1+\left(\frac{3}{\zeta}+\frac{3}{\zeta^2}\right)J_2+\left(-\frac{1}{\zeta}-\frac{1}{2\zeta^2}\right)J_3\right]
\label{LGZ}
\ee

\begin{figure}[H]
\centering
\includegraphics[scale=0.25]{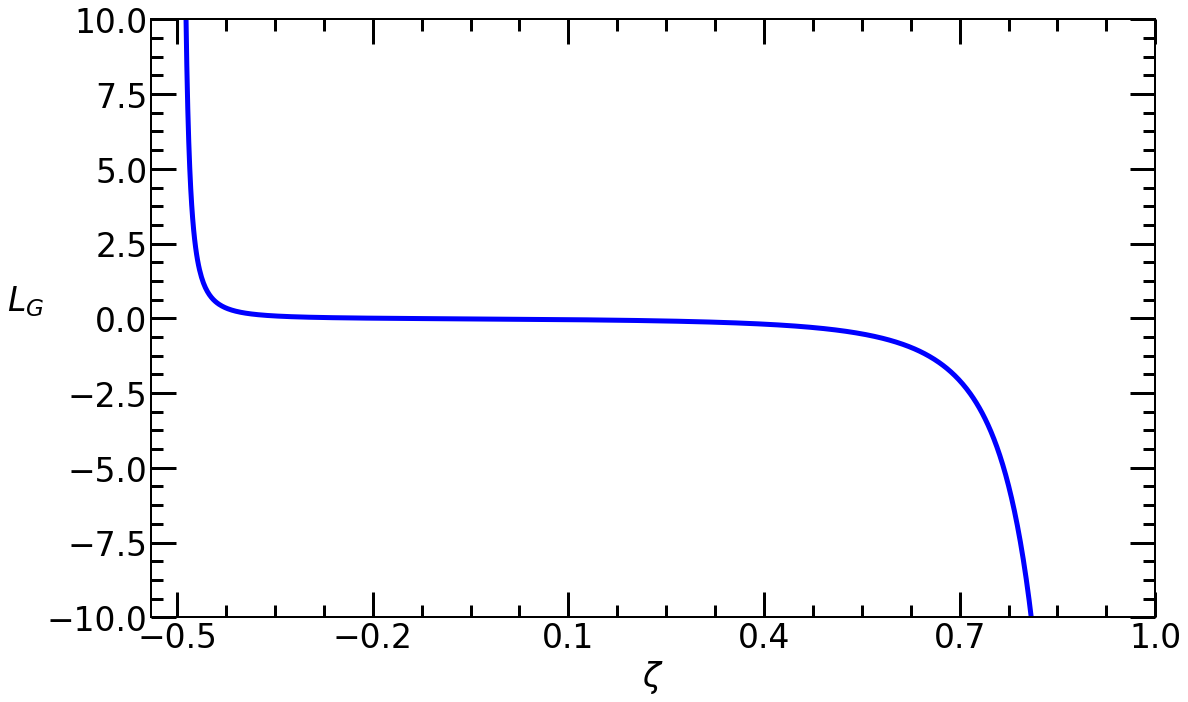}
\caption{$L_G$ function (see eq. (\ref{LGZ})) vs $\zeta$.}
\label{LGub}
\end{figure}

This function is plotted in Fig. \ref{LGub} vs $\zeta$ in natural scale from $\zeta=-1/2$ to $\zeta=1$. $L_G(\zeta)$ qualitatively has the same behaviour of $L(\zeta)$ for Bjorken (\ref{LZB}).

$\Phi^{\mu}$ (eq. (\ref{gen2})) may be written in terms of the $J_2$ function as in eq. (\ref{phiJ2}), and then
\bea
\label{entropyG}
s&=&\frac{T^3}{2\pi^2}J_2+\frac{1}{T}T^{\rho\rho}-\frac{\zeta}{T}(T^\theta_\theta+T^\phi_\phi-2T^\eta_\eta)\nn
&=&\frac{T^3}{2\pi^2}J_2+\frac{1}{T}\left[T^{\rho\rho}(1-\zeta)+3\zeta T^\eta_\eta\right]\nn
&=&\frac{2T^3}{\pi^2}J_2
\tea

\section{DTTs as second order theories}
In this appendix we shall compare DTTs to the better known so-called ``second order'' hydrodynamic theories, taking references \cite{DMNR10,DMNR11,DMNR11b,DMNR12,DMNR12b,DMNR14,DMNR12c,DMNR14b} and \cite{JRS14,FJMRS15,TJR17,TVNH19} as representative formulations.

For simplicity, we shall restrict our discussion to particles in Minkowsky space-time in the relaxation time approximation. The state of the fluid is described by a 1pdf $f$ obeying the Boltzmann equation (\ref{Boltzmann}) with collision integral (\ref{AW}). The energy-momentum tensor $T^{\mu\nu}$ is defined as in (\ref{EMT}) and is conserved. While not the only choice, for conformal particles, where there are no other conserved currents, it is convenient to follow Landau-Lifshitz in defining the fluid as the (only) time-like unit eigenvector of $T^{\mu\nu}$. We then obtain the decomposition (\ref{decomp}), with $T_0^{\mu\nu}$ given by (\ref{idealEMT}) and $\Pi^{\mu\nu}$ as in (\ref{viscousEMT}). The problem is that the four conservation equations (\ref{7}) are not enough to determine the nine independent components of $T^{\mu\nu}$. 

One possible solution is to simply provide a suitable expression for $\Pi^{\mu\nu}$ in terms of the already defined temperature and velocity, and their derivatives \cite{ChapmanCowling}. If restricted to first spatial derivatives, this leads to the so-called first order theories, with well documented stability and causality problems \cite{HL83,HL85,HL88a,HL88b,Ols90,HP01,DKKM08,PKR10,GPRR19}. These may be overcome by including higher derivatives \cite{BRSSS08,BHMR08}. Eventually we may include time derivatives of $\Pi^{\mu\nu}$ itself, whereby the supplementary expression actually becomes a dynamical equation for the viscous EMT, and leading us into the so-called ``second order'' theories.

Again the simplest approach would be to take a time derivative of (\ref{viscousEMT}), and then use the Boltzmann equation for $\dot f$ \cite{DMNR10}. This leads to a closure problem, because, for collision terms more realistic than the Anderson-Witting one, the resulting expression cannot be readily written in terms of $T$, $u^{\mu}$ and $\Pi^{\mu\nu}$ themselves.

The most common approach \cite{DMNR12,DMNR12b} starts instead for a parametrization of the 1pdf

\be
f=f_0\left(1+\delta f\right)
\label{param}
\te 
where $f_0=e^{\beta_{\mu}p^{\mu}}$ is a local equilibrium distribution function and $\delta f$ is expressed in terms of a set $N_{\alpha}$ of known functions of momentum, with space-time dependent coefficients

\be
\delta f=\sum_{\alpha}C_{\alpha}\left(x,t\right)N_{\alpha}\left(p\right)
\label{param2}
\te 
Quantum statistical effects are easily included and we shall not discuss them. Also we may restrict the $N_{\alpha}$ to the set of irreducible tensors. Finally, we may generalize this approach by allowing a more general type of base 1pdf $f_0$ \cite{BHS14,BNR16a,BNR16b,NMR17}, for example, to take $f_0$ within the Romatschke-Strickland class \cite{RS03}. For simplicity we shall not discuss these extensions of the basic theory.

The next step is to choose a second class of functions $M_{\beta}$ to form weighted averages of the Boltzmann equation

\be
\int\;Dp\;M_{\beta}\left(p\right)\left[p^{\mu}\partial_{\mu}f_0\left(1+\delta f\right)-I_{col}\left[f_0\left(1+\delta f\right)\right]\right]=0
\label{momeqs}
\te
Choosing a suitable number of momenta eq. (\ref{momeqs}) one should obtain evolution equations for all parameters $C_{\alpha}$ in eq. (\ref{param2}). To evaluate the relative importance of the different terms in eqs. (\ref{param2}) and (\ref{momeqs}), references \cite{DMNR12,DMNR12b} propose using an expansion in inverse Reynolds $\mathbf{Re}$ and Knudsen $\mathbf{Kn}$ numbers. The Reynolds number is defined as the ratio of a typical component of $T_0^{\mu\nu}$ to a typical component of $\Pi^{\mu\nu}$, while the Knudsen is the ratio of the microscopic and macroscopic lenght scales. 

Not far from equilibrium, we expect that a typical component of $T_0^{\mu\nu}$ will be of the order of the energy density $\epsilon$, while a typical component of the viscous EMT will be of the order of the shear viscosity times the shear tensor $\left|\Pi^{\mu\nu}\right|\approx\eta\left|\sigma^{\mu\nu}\right|$. Since also $\eta\approx \tau\epsilon$, where $\tau$ is the collision time, we get $\mathbf{Re}^{-1}\approx\tau\left|\sigma^{\mu\nu}\right|$. Similarly, we may take $\tau$ itself as a characteristic microscopic scale \cite{DMNR11b}, and the inverse of a typical component of the shear tensor as a characteristic macroscopic scale \cite{DMNR16}, thus arriving to the estimate $\mathbf{Kn}\approx\tau\left|\sigma^{\mu\nu}\right|\approx\mathbf{Re}^{-1}$ \cite{DMNR14b}. See \cite {DMNR14} for exceptions to this rule of thumb.

The allowed choices of functions $N_{\alpha}$ and $M_{\beta}$ are restricted by the requirement of nonnegative entropy production. Just as an example, let us consider a case where there is a single $N$ function

\be 
\delta f= C_{\mu\nu}\left(-u^{\mu}p_{\mu}\right)^Np^{\mu}p^{\nu}
\label{fo}
\te
$C_{\mu\nu}$ is assumed to be traceless and transverse. Then

\bea
\epsilon&=&\frac3{\pi^2}T^4\nn
\Pi^{\nu\rho}&=&\frac1{15\pi^2}T^{N+6}\Gamma\left[N+6\right]C^{\nu\rho}
\tea
We may elliminate $C^{\nu\rho}$ to get 

\be 
f=f_0\left(1+\frac{15\pi^2}{T^{-\left(N+6\right)}\Gamma\left[N+6\right]}\Pi_{\mu\nu}\left(-u^{\mu}p_{\mu}\right)^Np^{\mu}p^{\nu}\right)
\te
Of course the conservation equations for the EMT hold as usual. We seek to close the system of equations by demanding that

\be
H^{\lambda\sigma}_{\mu\nu}\int Dp\;\left(-u^{\mu}p_{\mu}\right)^Mp^{\mu}p^{\nu}\left[p^{\rho}f_{,\rho}+\frac {\left(-u^{\mu}p_{\mu}\right)}{\tau}\left(f-f_0\right)\right]=0
\te
The projector $H^{\lambda\sigma}_{\mu\nu}$ is defined in eq. (\ref{hproj}). Computing the integrals over momentum space we get

\bea
0&=&H^{\mu\nu}_{\lambda\tau}\partial_{\rho}\left[u^{\rho}\frac{\Gamma\left[N+M+7\right]}{\Gamma\left[N+6\right]}T^{M+1}\Pi^{\lambda\tau}+\frac1{2\pi^2}T^{M+5}\Gamma\left[M+5\right]\frac23u^{\lambda}\Delta^{\tau\rho}\right]\nn
&+&\frac M{15\pi^2}T^{M+5}\Gamma\left[M+5\right]\sigma^{\mu\nu}+\frac M{7}T^{M+1}\frac{\Gamma\left[N+M+7\right]}{\Gamma\left[N+6\right]}\nn
&&\left[u^{\lambda}_{,\lambda}\Pi^{\mu\nu}+\Delta^{\mu\rho}u_{\lambda,\rho}\Pi^{\lambda\nu}+\Delta^{\nu\rho}u_{\lambda,\rho}\Pi^{\lambda\mu}+u^{\mu}_{,\lambda}\Pi^{\lambda\nu}+u^{\nu}_{,\lambda}\Pi^{\lambda\mu}-\frac43\Delta^{\mu\nu}\Pi^{\lambda\sigma}u_{\lambda,\sigma}\right]\nn
&+&\frac{\Gamma\left[N+M+7\right]}{\tau\Gamma\left[N+6\right]}T^{M+1}\Pi^{\mu\nu}
\tea
or else, keeping only the second order terms

\be 
0=\frac{\Gamma\left[N+M+7\right]}{\Gamma\left[N+6\right]}\left[H^{\mu\nu}_{\lambda\tau}\dot\Pi^{\lambda\tau}+\frac1{\tau}\Pi^{\mu\nu}\right]+\frac 1{15\pi^2}T^{4}\Gamma\left[M+6\right]\sigma^{\mu\nu}
\te
corresponding to a shear viscosity
\be
\eta_{N,M} =\frac{\Gamma\left[M+6\right]\Gamma\left[N+6\right]}{15\pi^2\Gamma\left[M+N+7\right]}\tau T^{4}
\te
As is well known, different choices lead to different transport coefficients. 

In this theory, the entropy current is

\bea
S^{\mu}&=&\int Dp\;p^{\mu}f_0\left(1+\delta f\right)\left[1-\beta_{\rho}p^{\rho}-\delta f+\frac12\left(\delta f\right)^2\right]\nn
&=&\frac 4{\pi^2}T^3u^{\mu}-\frac12\int Dp\;p^{\mu}f_0\left(\delta f\right)^2
\tea
Now

\be
\int Dp\;p^{\mu}f_0\left(\delta f\right)^2=\frac{15\pi^2\Gamma\left[2N+7\right]}{T^{5}\Gamma\left[N+6\right]^2}\Pi_{\nu\rho}\Pi^{\nu\rho}u^{\mu}
\te
and 

\be
S^{\mu}_{,\mu}=-T^{-1}\Pi^{\rho\sigma}u_{\rho,\sigma}-\frac{15\pi^2\Gamma\left[2N+7\right]}{T^{5}\Gamma\left[N+6\right]^2}\Pi_{\mu\nu}\dot\Pi^{\mu\nu}-\frac12\frac{15\pi^2\Gamma\left[2N+7\right]}{T^{5}\Gamma\left[N+6\right]^2}\Pi_{\nu\rho}\Pi^{\nu\rho}\left[u^{\mu}_{,\mu}-5\frac{\dot T}T\right]
\te
We use the equation for $\Pi^{\mu\nu}$, discarding third order terms, to get

\be
S^{\mu}_{,\mu}=\frac{15\pi^2\Gamma\left[2N+7\right]}{\tau T^{5}\Gamma\left[N+6\right]^2}\Pi_{\mu\nu}\Pi^{\mu\nu}+\frac1T\left[\frac{\Gamma\left[2N+7\right]\Gamma\left[M+6\right]}{\Gamma\left[N+6\right]\Gamma\left[M+N+7\right]}-1\right]\Pi^{\rho\sigma}u_{\rho,\sigma}
\te
To enforce nonnegative entropy production, the last term must vanish. We see that, while this condition does not fix $N$ and $M$ uniquely, it does restrict the allowed choices. Most importantly, we see that the validity of the Second Law in hydrodynamics does not follow automatically from the kinetic theory $H$ theorem.

Unlike second order theories, DTTs start from an exponential ansatz

\be 
f=f_0e^{\bar\delta f}
\te
If, for example, we choose

\be 
\bar\delta f=T^{-\left(N+2\right)}\zeta_{\mu\nu}\left(-u^{\mu}p_{\mu}\right)^Np^{\mu}p^{\nu}
\te
(the parameter $T$ is no longer the Landau-Lifshitz temperature, since $T^{00}$ gets $\zeta$-dependent corrections), then we close the system of equations by asking that 

\be
H^{\lambda\sigma}_{\mu\nu}\int Dp\;\left(-u^{\mu}p_{\mu}\right)^Np^{\mu}p^{\nu}\left[p^{\rho}f_{,\rho}+\frac {\left(-u^{\mu}p_{\mu}\right)}{\tau}\left(f-f_0\right)\right]=0
\te
The resulting theory has nonnegative entropy production, as discussed in the main text. 

In order to compare this DTT to a generic second order theory, we must truncate the former to a finite order in inverse Reynolds and Knudsen numbers. It is easy to see that not far from equilibrium $\mathbf{Kn}\approx\mathbf{Re}^{-1}\approx\left|\zeta^{\mu\nu}\right|$. Thus, if we wish to keep terms up to second inverse Reynolds number, as in the second order theory we showed above, then we must expand the exponential

\be 
f\approx f_0\left(1+\bar\delta f+\frac12\left(\bar\delta f\right)^2\right)
\te
We conclude that a truncated DTT is a particular case within the class of theories discussed in \cite{DMNR12,DMNR12b} , albeit a very special one. What makes it special is that new terms are included into the 1pdf without enlarging the set of free parameters, and most importantly, enforcing nonnegative entropy production all along. The aim of DTTs is thus to obtain an acceptable macroscopic description, complying with the basic laws of energy-momentum conservation and nonnegative entropy production, while keeping the set of free parameters to an absolute minimum. Of course, this entails a loss of generality, but a definite gain in simplicity and predictive power.

\end{document}